\documentclass[sigconf]{acmart}
\renewcommand\footnotetextcopyrightpermission[1]{} % removes footnote with conference information in first column
\makeatletter
\renewcommand\@formatdoi[1]{\ignorespaces}
\makeatother

\usepackage{amsmath}
\usepackage{amsfonts}
\usepackage[linesnumbered,ruled,vlined]{algorithm2e}
\usepackage{amsthm}
\usepackage{caption}
\usepackage{subcaption}

\newcommand{\argmin}{\arg\!\min}
\newcommand{\floor}[1]{\left\lfloor #1 \right\rfloor}

\AtBeginDocument{%
  \providecommand\BibTeX{{%
    \normalfont B\kern-0.5em{\scshape i\kern-0.25em b}\kern-0.8em\TeX}}}

\acmConference[MobiCom '20]{The 26th Annual International Conference on Mobile Computing and Networking}{21-25 Sep, 2020}{London, United Kingdom}
\author{Jun Liu}
\orcid{0000-0003-0107-0636}
\email{jun.liu@student.unsw.edu.au}
\affiliation{%
	\institution{University of New South Wales}
	\city{Sydney}
	\state{NSW}
	\country{Australia}
	\postcode{2052}
}

\author{Weitao Xu}
\email{weitaoxu@cityu.edu.hk}
\affiliation{%
	\institution{City University of Hong Kong}
	\city{Hong Kong}
	\state{}
	\country{}
	\postcode{}
}

\author{Sanjay Jha}
\email{sanjay.jha@unsw.edu.au}
\affiliation{%
	\institution{University of New South Wales}
	\city{Sydney}
	\state{NSW}
	\country{Australia}
	\postcode{2052}
}

\author{Wen Hu}
\email{wen.hu@unsw.edu.au}
\affiliation{%
	\institution{University of New South Wales}
	\city{Sydney}
	\state{NSW}
	\country{Australia}
	\postcode{2052}
}

\begin{document}

\title[Nephalai]{Nephalai: Towards LPWAN C-RAN with Physical Layer Compression}


% \renewcommand{\shortauthors}{X et al.}

\renewcommand{\shortauthors}{}

\begin{abstract}
    We propose \emph{Nephalai}, a Compressive Sensing-based Cloud Radio Access Network (C-RAN), to reduce the uplink bit rate of the physical layer (PHY) between the gateways and the cloud server for multi-channel LPWANs. 
Recent research shows that single-channel LPWANs suffer from scalability issues. While multiple channels improve these issues, data transmission is expensive.
Furthermore, recent research has shown that jointly decoding raw physical layers that are offloaded by LPWAN gateways in the cloud can improve the signal-to-noise ratio (SNR) of week radio signals.
%handle weak radio signals by combining coherent frames. 
However, when it comes to multiple channels, this approach requires high bandwidth of network infrastructure to transport a large amount of PHY samples from gateways to the cloud server, which results in network congestion and high cost due to Internet data usage. 
In order to reduce the operation's bandwidth, we propose a novel LPWAN packet acquisition mechanism based on Compressive Sensing with a \emph{custom design dictionary that exploits the structure of LPWAN packets}, reduces the bit rate of samples on each gateway, and demodulates PHY in the cloud with (joint) sparse approximation. Moreover, we propose an adaptive compression method that takes the Spreading Factor (SF) and SNR into account. Our empirical evaluation shows that up to 93.7\% PHY samples can be reduced by \emph{Nephalai} when $SF = 9$ and SNR is high without degradation in the packet reception rate (PRR). With four gateways, 1.7x PRR can be achieved with 87.5\% PHY
samples compressed, which can extend the battery lifetime of embedded IoT devices to 1.7.

% We propose \emph{Nephalai}, a Compressive Sensing-based Cloud Radio Access Network (C-RAN), to reduce the uplink bitrate of the physical layer (PHY) between the gateways and the cloud server for multi-channel LPWANs. Recent research has shown that jointly decoding raw physical layer that are offloaded by LPWAN gateways in the cloud can handle weak radio signals by combining coherent frames. However, when it comes to multiple channels, this approach requires high bandwidth of network infrastructure to transport a large amount of PHY samples from gateways to the cloud server, which results in network congestion and high financial cost due to the Internet data usage. In order to reduce the bandwidth usage of this data offloading operation, we propose a novel LPWAN packet acquisition mechanism based on Compressive Sensing with a \emph{custom design dictionary to exploit the structure of LPWAN packets}, to reduce the bit rate of samples on each gateway, and to demodulate PHY in the cloud with (joint) sparse approximation. Moreover, we propose an adaptive compression method that takes Spreading Factor (SF) and Signal to Noise ratio (SNR) into consideration. Our empirical evaluation shows that up to 93.7\% PHY samples can be reduced by \emph{Nephalai} when $SF = 9$ and SNR is high without degradation in packet reception rate (PRR). With four gateways, 1.7x Packet Reception Rate (PRR) can be achieved with 87.5\% PHY
% samples compression, which can extend the battery lifetime of embedded IoT devices by 1.7 times.
\end{abstract}

\makeatletter
\def\blfootnote{\gdef\@thefnmark{}\@footnotetext}
\makeatother

\maketitle
\blfootnote{Preprint version, revised on Oct. 11, 2020}
\section{Introduction}

\label{section:introduction}

Low-Power Wide Area Networks (LPWANs) are emerging wireless technologies with features such as comprehensive signal coverage, low bandwidth, potentially small packet sizes, and long battery life \cite{Farrell2018}. One of the representatives is LoRa, which has been widely used in commercial and industrial applications, such as logistical tracking, smart agriculture and intelligent building\cite{sinha2017survey}.

LoRaWAN is a recognized MAC-layer LoRa protocol for reliable data transfer, and it is generally deployed on unlicensed ISM bands with 125 kHz or 500 kHz narrow band channels. Such narrow bands limit the bit rate down to several kilo-bits or hundred-bits per second, while they benefit the demodulator's sensitivity, making it possible to detect and decode LoRa signals significantly lower than noise floor.

Previous research demonstrates that if only one channel is used, LoRaWAN coverage drops exponentially as the number of end-devices grows \cite{Georgiou2017} and may only support approximately 120 nodes for a typical smart city deployment \cite{Bor}. Some other research similarly indicates that LoRaWAN can support from 200-1000 nodes in different applications~\cite{Liando2019,Xu2019}, which raises concerns about the scalability of LoRaWAN.  
To this end, by extending from single to multiple channels similar to frequency division multiple access (FDMA), the scalability can be increased~\cite{Liando2019}. 
Typical LoRaWAN gateways equipped with Semtech SX1301 chips \footnote{SX1301 datasheet.
%Hyperlink is removed according to the submission instruction.
~\url{https://www.semtech.com/products/wireless-rf/lora-gateways/sx1301}
}
can operate with up to 8 $\times$ 125kHz channels, which provides greater network capacity than a single channel network by eight times. 
% ===== revision =====
Furthermore, in the USA, up to 64 $\times$ 125kHz narrow-band channels are allocated on unlicensed ISM bands for LoRaWAN. A naive approach to cover more than eight channels is to use several gateways simultaneously in one spot. A commercial outdoor LoRaWAN gateway costs approximately US\$1,000. Therefore, covering all 64 channels would be expensive and difficult to maintain.
% ===== revision =====

Beyene et al. propose the implementation of NB-IoT via Cloud-Radio Access Networks, which are easy to implement and cost-efficient to deploy~\cite{Beyene2017}. NB-IoT and LoRa/LoRaWAN are both LPWAN technologies and share many common features. Inspired by the C-RAN of NB-IoT, we propose a C-RAN architecture for LoRaWAN as an affordable solution to support as many LoRaWAN channels as possible. Thus, with the help of software-defined ratios (SDR), parallel gateways are replaced with a single remote radio head, and PHY processing is offloaded to the cloud.

As an extra benefit of C-RAN, the opportunity to increase the battery life for end devices is provided. Some other approaches such as optimal frequency selection~\cite{gadre2020frequency} and backscatter~\cite{Talla2017} have been proposed, while our approach is based on spatial diversity gains. Similar to the architecture of cellular networks~\cite{Checko2015}, multiple LoRaWAN gateways are commonly deployed to provide wide-area network coverage. Therefore, the signal from one end device can be received by multiple gateways and processed jointly. In a recent research, Dongare et al. implemented such a system to exploit the spatial diversity gain to improve SNR by coherently combining PHY samples captured by various gateways in different locations~\cite{Dongare2018}. Thus, an end-device may transmit with a faster bit rate, which results in a shorter transmission duration for a fixed packet/data payload length. Their evaluation shows that increasing the number of received gateways improves the SNR of packets in an approximately logarithmic manner.

Although the aforementioned C-RAN is a promising architecture with many benefits for IoT wireless networks,
such a system has a huge impact on the PHY offloading network between the gateways and the cloud. According to Charm~\cite{Dongare2018}, when a moving average compressed technique is applied for PHY, 9 Mbps is required for each 500kHz channel and 2.25 Mbps is for each 125kHz channel respectively, which produces $2.25$ Mbps $\times$ 64 = 144 Mbps data traffic to the cloud if a gateway supports 64 $\times$ 125kHz LoRa channels. For lossless Nyquist sampling and data stored as 24-bit I/Q samples (12-bit for I/Q each, same as SX1301), a minimal bit rate of $24$ bit$ \times (64 \times 125$kHz$) = 192$ Mbps is required for the PHY offloading network. Both settings require gigabit bandwidth for reliable data transmissions, which is challenging in both outdoor or indoor scenarios such as pastures and buildings with sub-100-megabit Internet connections. 
Moreover, in some rural areas, Internet can only be provided via satellites, the bandwidth of which is very limited. On the other hand, a large-scale LoRaWAN (e.g., with hundreds of gateways) will pose a significant traffic to the data center. It may influence the real-time delivery of PHY samples and reduce the performance of joint decoding that requires
synchronized PHY samples from different gateways. 

One solution is to equip optical fibers as part of the infrastructure of the PHY dispatching network. However, the cost is unaffordable for many low-cost or ad hoc IoT applications. Another solution is to upload active channels only. However, for large-scale deployment (i.e., tens of thousands of nodes), the probability of simultaneous multi-channel occupation
is high. Moreover, because low SNR signals can benefit from joint processing in the cloud, the channel activity detector becomes more sensitive and uploads PHY samples of idle channels to the cloud due to ``false alarms''. 

\begin{figure}[tb]
	\centering
	\includegraphics[width=0.78\linewidth]{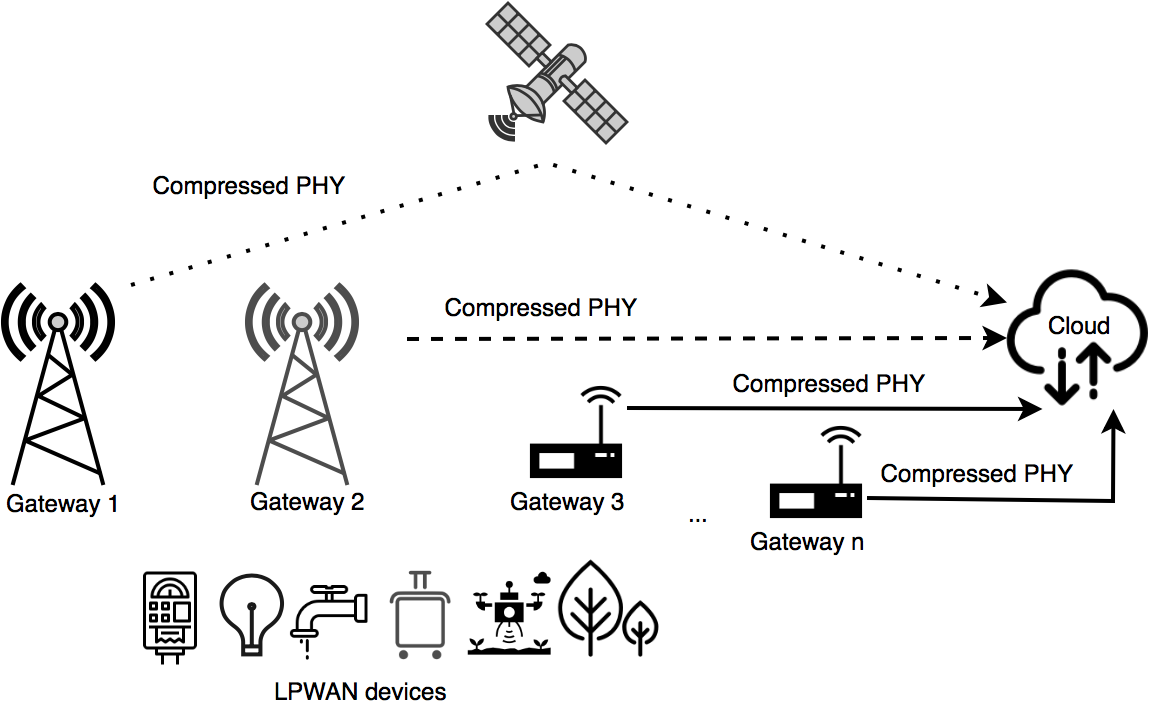}
	\caption{The overview of \emph{Nephalai} decoding in the cloud with compressed PHY samples.}
	%(\textcolor{red}{This graph is too detailed. You'd better give an overview like Fig.1 in Charm first and then plot a graph like Fig. 3 in Charm.})}
	\label{fig:system-overview}
% 	\vspace{-10pt}
\end{figure}

Therefore, PHY compression is the key enabler for LPWAN C-RAN. To this end, we propose a Compressive Sensing (CS)-based technique, called \emph{Nephalai}\footnote{In ancient Greek mythology, \emph{Nephalai} is the nymph of the clouds.}, to reduce the network bandwidth between gateways and the cloud. 
Fig.~\ref{fig:system-overview} shows the overview of \emph{Nephalai}, which leverages the sparsity of the PHY for signal compression and (joint) reconstruction. 
Dictionaries and measurement matrices in \emph{Nephalai} are custom-designed to \emph{exploit the structure of LoRa radio signals} to achieve the best compression and reconstruction performance. 
\emph{Nephalai} is designed to run in real-time and is implemented with SDR  {\footnote{One limitation for \emph{Nephalai} is the front-end hardware. Although our prototype discussed in Sec.~\ref{section:architecture} later can support 64 channels, if \emph{Nephalai} is implemented on legacy front-end SX1257, it can support 8 channels only.}}.
Our testbed evaluation in our campus has shown that, 1) up to 93.7\% samples can be reduced without packet reception rate (PRR) reduction; 2) \emph{Nephalai} can improve battery lifetimes to 1.7x with four gateways and 87.5\% PHY samples compressed.

The contributions of this paper are as follows.

\begin{itemize}
	
	\item We propose a novel CS-based compression technique for cloud-assisted LPWAN that significantly reduces the bandwidth between the gateways and the cloud. 
	
	\item We propose a new dictionary to achieve high compression ratios without performance degradation. The proposed dictionary 
	exploits the structure of LoRa radio signals, and achieves more than two orders-of-magnitude better sparse representation than standard Discrete Fourier transform (DFT) and Discrete Cosine transform (DCT) domains.

	\item We implement a prototype of \emph{Nephalai} with software-defined radios, and
	our empirical evaluation demonstrates its superior 
	performance on embedded devices.
\end{itemize}

\section{Related work}
\label{section:background}

\subsection{LPWAN, LoRa, LoRaWAN}

%\textcolor{blue}{cite the sota lora demodulation/decoding research, open-source software}

LPWAN~\cite{Farrell2018,sinha2017survey, centenaro2016long} has attracted much attention from both academia and industry in recent years. 
%NB-IoT~\cite{ratasuk2016nb,Beyene2017,wang2017primer} has been standardized for cellular station to communicate with LPWAN Internet of Things (IoT) devices over licensed spectrum, while 
LoRaWAN~\cite{sinha2017survey,Adelantado2017,Saari2018} is standardized by the LoRa Alliance for LPWAN on an unlicensed spectrum.
LoRa~\cite{seller2016low,Knight2016,robyns2018multi,ghanaatian2019lora,seller2018low,Vangelista2015} is the physical-layer foundation of LoRaWAN and defines modulation and radio communication. 
Recent research proposes slotted ALOHA based on a synchronization technique~\cite{Polonelli2019}, which inspires us to synchronize LoRa symbols based on a similar scheme.
In this paper, we focus on the sparsity of LoRa signals, and leverage the structure of LoRa in demodulation to optimize the performance of physical-layer compression.

\subsection{C-RAN}
\label{section:relatedwork-cran}
%\textcolor{blue}{title may change to LoRa in the cloud, adding LoRaWAN geolocation white paper, by looking into physical layer, the sensitivity is increased for low SNR, TDOA calculation is more accurate...}
C-RAN was proposed originally for cellular networks based on the concepts of centralization and virtualization for the baseband operations~\cite{ChinaMobile2011, Checko2015, Wubben2014, Beyene2017}. The network can process demodulation
%\footnote{We use modulation and encoding, and demodulation and decoding interchangeably in this paper.} 
in the cloud coherently, to exploit the diversity scheme. This refers to improving the reliability of wireless communication by using multiple radio channels~\cite{Brennan1959}. 
%Such technique of cloud-assisted coherently decoding in physical layers has 
Such cloud-assisted decoding techniques in physical layers have also been investigated in Wi-Fi \cite{Tan2009,Xie2014} and LPWAN \cite{Dongare2018,hoeller2018analysis,Beyene2017}. 
%Particularly, a recent C-RAN system for LPWAN called Charm~\cite{Dongare2018}, exploits spatial diversity to increase battery life for end devices.
%improve LPWAN decoding by coherently combining signals captured by various gateways in different locations. 
% The SNR of the combined signal is improved, so the embedded end-devices may select higher data rates for transmissions which results in shorter transmission duration and improves the battery life of the devices. 
The modification of gateways is transparent to the senders; therefore, there is no requirement for changes to the original embedded LoRa devices, which maintain their compatibility with the legacy devices.

\subsection{Physical Layer Compression}
%\textcolor{blue}{cite how PHY compression is done for other wireless communication technologies. }
One challenge of C-RAN is the high bandwidth requirement in transmitting I/Q streams from the edges to the cloud~\cite{Checko2015, Beyene2017}. 
%A typical scenario of LTE C-RAN for 20 MHz band and 16-bit I/Q samples with $2 \times 2$ MIMO has 2.5 Gbps bit rate in front-haul~\cite{Checko2015}. 
The Wyner-Ziv coding scheme leverages the correlation (side information) among receivers to use a finer quantizer in PHY compression \cite{Xia2018,Park2014}.
%so to reduce the required front-haul transmission rate. 
However, the implementation of a distributed Wyner-Ziv compression is challenging mainly due to the complexity of obtaining the optimal joint compression codebook and the joint decompressing/decoding in the cloud~\cite{Peng2015}.
% Wyner-Ziv compression is difficult mainly due to the high complexity in determining the optimal joint compression codebook and the joint decompressing/decoding at the BBU pool
%\cite{wang2015comressive} CS
Alternatively, Compressive Sensing has been applied to achieve distributed front-haul compression \cite{Rao2015,wang2015compressive}. However, designing a sparse representation exploiting LoRa structures to improve compression performance has not yet been studied.
To this end, the proposed custom-designed dictionary and measurement matrices achieve
more than two orders-of-magnitude better performance than conventional DCT and DFT domains
used in prior work.
%One may raise the question, is it necessary to compress LPWAN physical layer?
%\textbf{The necessity of physical layer compression for LPWAN.} Although narrow band LPWAN C-RAN has lower bit rate requirement for fronthaul~\cite{Beyene2017}, the example discussed in Sec.~\ref{section:introduction} requires 36 Mbps, which is more than the typical broadband Internet bandwidth\footnote{Broadband benchmark for downloading is 25 Mbps according to FCC.  \url{https://www.fcc.gov/reports-research/reports/broadband-progress-reports/2015-broadband-progress-report}}. LoRaWAN with up to 20 MHz spectrum composed of 96 consecutive channels can generate 640 Mbps bit rate for 16-bit I/Q samples, which is a huge burden for the network infrastructure, let alone simultaneous transmission by multiple gateways. Moreover, the over-sampling technique that increases SNR~\cite{Gray1990} and reduces SER (see Fig.~\ref{fig:eva_oversampling} later) can make the fronthaul bit rate much higher.
%To this end, we propose a C-RAN architecture using CS theory for LoRa physical layer compression to address the problem. 

\textbf{Summary}: \emph{Nephalai} is partly inspired by Charm~\cite{Dongare2018}, but makes significant contributions towards reducing the traffic between LoRaWAN gateways and the cloud. Charm focuses on improving SNR and battery life with multiple gateways, while our work \emph{Nephalai} focuses on I/Q compression to further increase the capacity for PHY processing in the cloud. The compression technique used by Charm is the sum of consecutive samples in windows and generates data at a rate of 9 Mbps for a 500 kHz band. Taking our evaluation in Sec.~\ref{subsubsec:singledecoding} as an example, with 87.5\% compression ratio, the data rate is 375 kbps for 125 kHz, equivalent to 1.5 Mbps for 500 kHz (more than 80\% reduction compared to Charm), which can help Charm further reduce the data rates between the gateways and the cloud. Thus, the proposed compression technique is complementary to Charm. Furthermore, the proposed compression technique 
%can not only benefit Charm, but 
can also be applied in other scenarios such as multiple (or full) channel reception.

\section{Background}

\label{section:lorapremier}

\subsection{LoRa physical layer}

LoRa uses chirp spread-spectrum (CSS) as the method for modulation ~\cite{seller2016low,Vangelista2017}. The Spreading Factor ($SF$) is usually defined as an integer from 7 to 12, representing the number of encoded bits per chirp symbol. Bandwidth ($BW$) is the spectrum constraint of a channel, typically 125 or 500 kHz~\cite{LoRaAlliance2017}. As discussed in Sec.~\ref{section:introduction}, the LoRaWAN gateway uses 125 kHz %instead of 500 kHz 
for receiving packets from end devices, and thus in this paper we only focus on 125 kHz channels. 
% (\textcolor{red}{why we only focus on 125KHZ in this paper, should give some reasons}). 
%$SF$ and $BW$ affect transmission time, receiver sensitivity, and data rate.
LoRa utilizes time-shifted chirps in symbol modulation to carry information.
% An up-chirp has its frequency linearly increasing, 
The frequency of an up-chirp increases in a linear manner, while a down-chirp is the opposite.

\subsection{Demodulation}

\label{section:demodulation}

The commonly used demodulation method is pulse compression, where the chirp symbol is first multiplied by a down-chirp in the time domain, and then processed with Fast Fourier transform (FFT) \cite{seller2016low,Knight2016}.
The result is indicated by the most significant component in the frequency domain. 
%An example is shown in Fig.~\ref{fig:demod_11} and~\ref{fig:demod_12} for well segmented or synchronized (see Sec.\ref{sec:sync}) chirps. 
If the symbol is not well segmented or unsynchronized, several peaks instead of one may show up,
%be seen as shown in Fig.~\ref{fig:demod_21} and~\ref{fig:demod_22}, 
which results in demodulation failure. 
%After demodulation, further decoding is required to recover the original bits that has been sent. 
Open-source software such as gr-lora \cite{Knight2016,Robyns2017} provide demodulation and decoding functions, while \emph{Nephalai} focuses on PHY compression only. 

% \vspace{-5pt}

% \begin{figure}[htb]
% 	\centering
% 	\begin{subfigure}[t]{0.49\linewidth}
% 		\centering
% 		\includegraphics[width=\linewidth]{figures_new/demod_11.png}
% 		\caption{Well segmented/sync}
% 		\label{fig:demod_11}
% 	\end{subfigure}
% 	\hfill
% 	\begin{subfigure}[t]{0.49\linewidth}
% 		\centering
% 		\includegraphics[width=\linewidth]{figures_new/demod_12.png}
% 		\caption{De-chirping of (a)}
% 		\label{fig:demod_12}
% 	\end{subfigure}
% 	\hfill
% 	\begin{subfigure}[t]{0.49\linewidth}
% 		\centering
% 		\includegraphics[width=\linewidth]{figures_new/demod_21.png}
% 		\caption{Badly segmented/unsync}
% 		\label{fig:demod_21}
% 	\end{subfigure}
% 	\hfill
% 	\begin{subfigure}[t]{0.49\linewidth}
% 		\centering
% 		\includegraphics[width=\linewidth]{figures_new/demod_22.png}
% 		\caption{De-chirping of (c)}
% 		\label{fig:demod_22}
% 	\end{subfigure}
% 	\vspace{-5pt}
%     \caption{Spectrogram of typical PHY. 
% 	%(a) Synchronized symbols aligned with de-chirping window. (b) One significant peak in frequency domain for symbols in aligned de-chirping window. (c) Symbols are misaligned with de-chirping window. (d) Multiple significant peaks of frequency domain for symbols in misaligned de-chirping window.
% 	}
% 	\vspace{-10pt}
% 	\label{fig:demodulation}
% \end{figure}

% \vspace{-8pt}

\subsection{Synchronized symbol}
\label{sec:sync}

\begin{figure}[htb]
	\centering
	\includegraphics[width=0.93\linewidth]{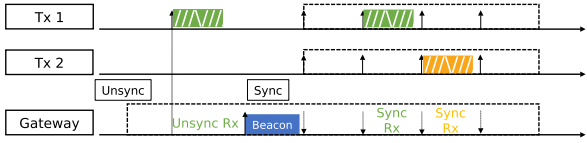}
	\caption{Synchronized receiving for chirp symbols}
	\label{fig:sync:rx}
\end{figure}

Inspired by LoRaWAN class B \cite{LoRaAlliance2017} and slotted ALOHA \cite{Polonelli2019}, we can synchronize end nodes and gateways so gateways can receive with non-overlapped windows as shown in Fig.~\ref{fig:sync:rx}. 
However, perfect synchronization is neither possible nor necessary. Here, we use synchronized reception to improve the compression performance only, and further digital signal processing is performed in the cloud for fine-grain symbol segmentation. Thus, the synchronization error tolerance is high. This will be discussed further in Sec.~\ref{subsec:symbolcompression}.

\vspace{-5pt}

\vspace{-8pt}

\section{Architecture}

\label{section:architecture}

The \emph{Nephalai} system has one cloud server equipped with GPU for
$\ell_1$ minimization acceleration, and inexpensive single-board computers with SDRs as the edge gateways. Physical-layer radio samples are transferred from gateways to the cloud server via conventional Internet infrastructure.% such as Wi-Fi and Ethernet. 

\begin{figure*}[hbt]	
	\centering
	\vspace{-3pt}
	\includegraphics[width=0.8\linewidth]{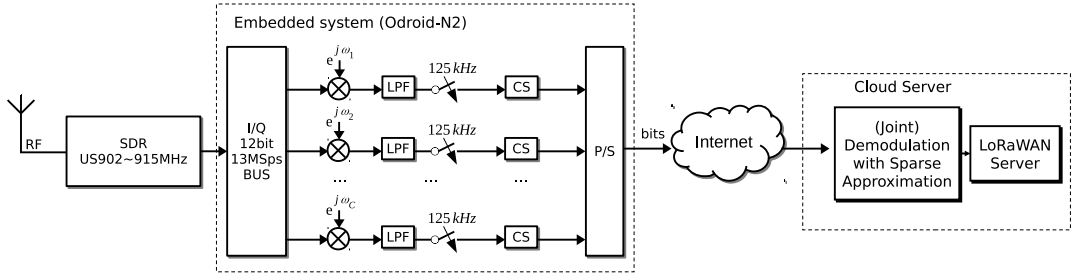}
	\vspace{-6pt}
	\caption{Baseband block diagram showing the architecture of \emph{Nephalai}}
	%(\textcolor{red}{This graph is too detailed. You'd better give an overview like Fig.1 in Charm first and then plot a graph like Fig. 3 in Charm.})}
	\label{fig:clora-architecture}
	\vspace{-5pt}
\end{figure*}

Fig.~\ref{fig:clora-architecture} depicts the overall architecture of \emph{Nephalai}.
%, which is low-cost and easy to deploy. 
The gateway clocks are synchronized via PPS from GPS modules 
with the accuracy of several microseconds. The accurate timestamp can help synchronize LoRa chirp symbols 
(see Fig.~\ref{fig:sync:rx}) 
and help the cloud server detect coherent LoRa packets easily. 
% As the typical duration of LoRa packet varies from tens of milliseconds to seconds, the error of several milliseconds is able for \emph{Nephalai} to distinguish incoherent LoRa packets. 
%GPS module and RTC clock (\textcolor{red}{what is RTC})may be attached for better synchronization. 
% Better time synchronization may be achieved via other methods such as GPS.
%\textcolor{red}{Complexity analysis goes here}
To analyze the complexity of our encoding algorithm in edge devices\footnote{We omit the complexity analysis of the proposed decoding algorithm in the cloud (i.e., $\ell_1$ minimization solver) since the cloud can be seen as having unlimited resources.}, suppose we have $N$ samples per symbol (this will be discussed in Sec.~\ref{subsubsec:Phi}, $N=128$ in practice), $M$ samples per compressed vector, $C$ as the number of channels and $P$ as the number of low pass filter (LPF) taps. Then, the frequency conversion block together with LPF is $O(NP)$, the down-sampler is $O(N)$, and the CS block is $O(MN)$. The overall complexity in the edge devices is $O(NPC+NC+NMC)$. Therefore, fewer taps for LPF and higher compression ratio for CS block can improve the performance of the embedded system. 
In order to support multiple 125 kHz channels as discussed in Sec.~\ref{section:introduction}, the SDR of the gateway captures the whole 13 MHz LoRa spectrum, and the embedded system filters each channel and compresses using a shared measurement matrix. Compressed bits of each channel are packed together and uploaded to the cloud server. The cloud server then performs decompression and demodulation to recover the LoRa chirp symbols or jointly process all coherent symbols to improve their accuracy. 
%The cloud server remotely controls the gateways with radio frequency, bandwidth and measurement matrix. 
%and thus the system can be easily configured to sense TV white spaces or deploy on licensed band with permit where conventional LoRaWAN gateways are not capable. 
 
% \vspace{-2pt}

\section{Compression}

\label{section:compression}

% \textcolor{blue}{Under re-construction}

%In this section, we discuss the algorithms used in LoRa physical layer compression.

\subsection{Lossless compression}
\label{subsec:gzip}
We compare the compression performance of \emph{Nephelai} against a conventional lossless compression LZ77-based algorithm, \emph{gzip} \cite{Misra2014}.
% (\textcolor{red}{need a reference for this method}).
Gzip can only achieve a 7.5\% compression ratio for Nyquist-sampled LoRa PHY, which means 92.5\% of samples are not compressible. Such a low compression ratio is due to the fact that chirps spread across the whole spectrum, and general compression algorithms cannot exploit this sparsity in the frequency domain. In the following discussion, we consider the lossless compression ratio as the baseline, and investigate a novel CS-based algorithm to increase the compression ratio.

\subsection{Compressive Sensing}
\label{subsec:cs}

CS is an information theory~\cite{Baraniuk2007, Donoho2006,Candes2006} that proposes an approach to recover high dimensional \emph{sparse} signals from low dimensional measurements. Table~\ref{table:terminology} summarizes the mathematical symbols in this discussion.
For a predefined dictionary $\Psi \in \mathbb{C} ^ {N \times D}$,
%each column $\mathbf{\psi}_i \in ~\mathbb{R} ^ {N}$  represents an basis of $\Psi$ domain, and 
any signal $\mathbf{x} \in \mathbb{C} ^ N$ can be a linear combination of $\Psi$ as:
\begin{equation}
\label{e1}
\mathbf{x} = \Psi \mathbf{s} % = \sum_{i=1}^{D} s^{(i)} \vec{\psi}_i
\end{equation} 

% \vspace{-3pt}

where $\mathbf{s}  \in \mathbb{C} ^ D$ is a coefficient vector of $\mathbf{x}$  in the $\Psi$	 domain. If $N < D$, given $\mathbf{x}$ and $\Psi$, we can not solve Eq.~(\ref{e1}) to
obtain $\mathbf{s}$ in a general form because it is an undetermined problem. 

CS imposes the requirement that vector $\mathbf{s}$ is sparse; namely,
most of the elements in $\mathbf{s}$ are zeros.
Let $K$ denote the number of
non-zeros in $\mathbf{s}$, then $\mathbf{s}$ is sparse if $K << D$. $K$ in CS is termed as \emph{sparsity}. CS theory states that vector $\mathbf{s}$ can be recovered accurately by solving the following \emph{stable} $\ell_1$ minimization problem:
\begin{equation} \label{eq:l1}
\hat{\mathbf{s} } = \arg\min \left\| \mathbf{s} \right\|_1 \quad  s.t. \quad  \left\| \mathbf{x} - \Psi \mathbf{s}  \right\|_2 < \epsilon
\end{equation} 

% \vspace{-4pt}

where $\epsilon$ is noise, and provided that $\Psi$ satisfies the Restricted Isometry
Property (RIP) condition.
Note that RIP is only a sufficient but not a necessary condition.
Therefore, $\ell_1$-minimization may still be able to recover
the sparse $\mathbf{s}$ accurately, even if $\Psi$ does
not satisfy RIP. 
In fact,
$\ell_1$ minimization  % citation of compressinve sensing, two classic Transactions on Information Theory 2006 papers
has a rich history as it has been used to efficiently obtain useful \emph{sparse} information in the signals from a compressed representation
\cite{logan1965properties, Misra2011}.

%In particular,
%Misra et. al. proposed an energy efficient computing framework for GPS acquisition based on CS in~\cite{Misra2014}. The motivation is to offload GPS computation from embedded GPS receivers to a powerful central cloud to reduce the power consumption of the receivers. They observed that the size of raw GPS radio samples are large due to
%high sampling rates (e.g., in MHz), but they consist of very small amount of (e.g., sparse) useful information (i.e., time of arrival) only, which can be exploited by CS efficiently. Specifically, a random Bernoulli measurement matrix, where
%the number of rows is significantly smaller than the number of columns, is applied to the raw GPS radio samples for compression, to reduce the communication bandwidth and/or storage of the GPS receivers. After the compressed GPS signal is received by the cloud, a sparse approximation algorithm based on $\ell_1$ minimization will be applied for signal reconstruction to obtain the useful information (e.g., time of arrival of the GPS signal).

Common $\ell_1$ minimization algorithms are Matching Pursuit (MP), Orthogonal Matching Pursuit (OMP), Homotopy, $\ell_1$-magic, etc., and
the reconstruction performance of the algorithms depends on the sparsity of the signal and the incoherence between the measurement (compression) matrix and the signal itself, which is application dependent. 
Therefore, \emph{Nephelai} uses a
custom-designed dictionary $\Psi$ to exploit the structure of LoRa signal and a custom-designed measurement matrix $\Phi$ to maximize the
incoherence between the matrix $\Phi$ and the dictionary $\Psi$. Furthermore, \emph{Nephelai} features a unique joint decoding process to exploit the spatial diversity of the LoRa signals received
by the gateways in different locations to further improve the signal reconstruction (i.e., the decoding of the LoRa packets) performance.

\begin{table}[tbh]
	\centering
	\caption{The summary of mathematical symbols used.}
    \vspace{-5pt}
	\label{table:terminology}
	\small{
		\begin{tabular}{cl}
			\hline
			\multicolumn{1}{l}{Symbol} & \multicolumn{1}{l}{Definition} \\ \hline
% 			SF & LoRa Spreading Factor\\
% 			BW & LoRa bandwidth \\
			%			$R_b$ & The bit rate of LoRa\\
			$\Psi$ & CS dictionary\\
			%			$\Psi_d$ & A dictionary without phase shift\\
			$\Phi$ & CS measurement matrix \\ 
			$U$ & Diagonal matrix for up-chirp  \\
			$F_s$ & Sampling rate  \\
			$T$ & LoRa symbol duration  \\
			%$\Phi_{S}$ & CS measurement matrix based on SVD of $\Psi$ \\ 
			%$\Phi_{G}$ & CS measurement matrix based on Gaussian   \\ 
			%$\Phi_{B}$ & \begin{tabular}[c]{@{}l@{}}CS measurement matrix based on balanced \\ symmetric ($\pm 1$) Bernoulli distribution \end{tabular}\\
			$\mathbf{x}$ & Raw samples before compression \\
			$\mathbf{y}$ & Compressed vector of measurement\\
			$\mathbf{s}$ & Sparse vector\\
			% $\mathbf{r}$ & Residual\\
			$\alpha$ & Compression ratio\\
			$K$& The degree of sparsity \\
			$D$& The number of items in dictionary\\
			$N$& The number of complex samples in LoRa symbol \\
			$M$ & The length of compressed vector $y$\\
			\hline
		\end{tabular}
	}
    \vspace{-5pt}
\end{table}

\subsection{Dimension Reduction}
Johnson-Lindenstrauss Lemma shows that random projections can preserve the $\ell_2$ distance of vector $\mathbf{x} \in \mathbb{C} ^ N$ in a compressed domain $\mathbf{y} \in \mathbb{C} ^ M$, where $M < N$ with a high probability~\cite{Baraniuk2007} as:
\vspace{-4pt}
\begin{equation}
	\label{eq:y}
	\mathbf{y} = \Phi \mathbf{x} = \Phi (\Psi \mathbf{s})
\end{equation}

where $\Phi \in \mathbb{C} ^ {M \times N}$ is a random compression matrix
(recall that $\mathbf{x} = \Psi \mathbf{s}$ from Eq.~(\ref{e1}), $\Psi \in \mathbb{C} ^ {N \times D}$). 
Since the sparsity of $\mathbf{s}$ is $K$ (see Sec.~\ref{subsec:cs}),  Wright et al. show
that the minimum dimension of $M$ for a successful $\ell_1$ 
minimization recovery in practice is~\cite{wright2009}:
\vspace{-4pt}
\begin{equation}
	\label{eq:M}
	M \geq 2 K log(D/K).
\end{equation}

Substituting Eq.~(\ref{eq:y}) to (\ref{eq:l1}), $\ell_1$ minimization can be used
to recover sparse vector $\mathbf{s}$ from compressed measurement $\mathbf{y}$ as:
\begin{equation} \label{eq:l1_new}
	\hat{\mathbf{s}} = \arg\min \left\| \mathbf{s} \right\|_1 \quad  s.t. \quad  \left\| \mathbf{y} - \Phi (\Psi \mathbf{s})  \right\|_2 < \epsilon.
\end{equation} 

Therefore, instead of uploading raw LoRa radio samples $\mathbf{x} \in \mathbb{R} ^ N $ to the cloud,
a \emph{Nephelai} edge gateway uploads compressed measurements $\mathbf{y} \in \mathbb{R} ^ M$,
and achieves a \textbf{compression ratio} of $\alpha$ as:
\vspace{-4pt}
\begin{equation}
	\alpha = 1-M \div N.
	\label{equ:alpha}
\end{equation}

\subsection{Physical layer Compression}
%\label{subsec:compressionfinesegmentedsymbols}
\label{subsec:symbolcompression}

LoRa gateways can compress physical layer radio samples with a predefined measurement matrix ($\Phi\in \mathbb{C}^{M \times N}$, where $M < N$) before transmitting the 
compressed samples ($\mathbf{y} \in \mathbb{C}^M$) to the cloud server, where (joint) demodulation is performed based on the compressed signals by solving an $\ell_1$ 
minimization problem, i.e., Eq. (\ref{eq:l1_new}). 

For $SF\in \{7,8,9,10\}$, we propose one dictionary for each $SF$ covering two scenarios: 1) synchronized chirp symbol; 2) unsynchronized chirp symbol. Generally, scenario 2 is more common, and scenario 1 can be considered as a special case of scenario 2. Thus, a dictionary for unsynchronized should also be feasible for synchronized chirp symbols. However, based on our simulation and evaluation (see Sec.s~\ref{subsec:dectionarydesign} and~\ref{sec:eva:compressionratio}), the compression ratio of the synchronized chirps is better than that of the unsynchronized, and thus we recommend the implementation of the synchronization mechanism for LoRaWAN to achieve a better compression performance.

\subsubsection{Dictionary Design}
%\label{subsec:full_recovery}
\label{subsec:dectionarydesign}
Rao et al. have proposed the \emph{continuous}, direct compression of physical layer radio samples with non-overlapped windows, in an attempt to fully recover the signal from the cloud~\cite{Rao2015}. Normally, radio signals are sparse and compressible in conventional domains such as DFT and DCT. 
%We use a fixed window size the same as the symbol size $N$, and design a dictionary based on DFT basis.
For LoRa, such methods are applicable but a more sparse domain can be obtained by exploiting the structure of the signals.

As discussed previously in Sec.~\ref{section:demodulation}, 
%Fig.~\ref{fig:demod_21} shows an unsynchronized scenario. 
% We produce Fig.~\ref{fig:demod_22} 
we demodulate the symbols by multiplying the symbols with an ideal down-chirp in the time domain and then by performing FFT on the de-chirped symbol.
%The spectrogram shows that even a general 
Both synchronized and unsynchronized blocks are sparse in frequency after being multiplied by a down-chirp. Here we define \emph{block} as any $T$-length clip of a LoRa PHY, where $T$ is equal to the duration of one chirp.
%while \emph{symbol} is synchronized and well segmented. 
A block is a combination of parts from two consecutive symbols. 
%(between the two dash-line as shown in Fig.~\ref{fig:demod_21} and Fig.~\ref{fig:demod_22}). 
In the following sections,  \emph{block} and \emph{unsynchronized symbols} are interchangeable.
% \textcolor{blue}{Prior work in C-RAN compression based on CS~\cite{}}

% First, by letting $\lambda=0$ in Eq. (\ref{eq:v}), 
First, by letting $\varphi(t)$ stand for the phase of an ideal up-chirp,
we define matrix $\mathbf{U}$ as having a diagonal made of an ideal down-chirp (opposite phase to an up-chirp),
\begin{equation}
\label{eq:dict:1}
\vspace{-1pt}
\mathbf{U} = diag( e^{-j\varphi (\frac{0}{BW})}, e^{-j\varphi (\frac{1}{BW})},...,e^{-j\varphi (\frac{2^{SF}-1}{BW})} )
\end{equation}

% \begin{equation}
% \label{eq:dict:1}
% \mathbf{U} = diag( e^{-j\varphi^{(0)} (\frac{0}{BW})}, e^{-j\varphi^{(0)} (\frac{1}{BW})},...,e^{-j\varphi^{(0)} (\frac{2^{SF}-1}{BW})} )
% \end{equation}

Second, we define $\mathbf{W}$ as the DFT matrix for $N=2^{SF}$,
\vspace{-1pt}
\begin{equation}
\label{eq:dict:2}
\mathbf{W} = (\frac{\omega^{ik}}{\sqrt{N}})_{i,k=0,...,N-1}
\end{equation}

where $\omega=e^{-2 \pi j/N}$. Therefore, we can write a sparse representation for any LoRa block $\mathbf{x}$ as,
\vspace{-1pt}
\begin{equation}
\label{eq:dict:3}
\mathbf{s} = \mathbf{W} \mathbf{U} \mathbf{x}
\end{equation}

where $\mathbf{s}$ represents the frequency domain and has only a few non-zeros. Comparing Eq. (\ref{eq:dict:3}) and (\ref{e1}), we can then derive the dictionary $\Psi$ as,
\vspace{-3pt}
\begin{equation}
\label{eq:psi}
\Psi = \mathbf{U}^{-1} \mathbf{W}^{-1}
\end{equation}

where $(.)^{-1}$ is the matrix inversion. Therefore, the dictionary based on the sparsity of LoRa chirps is generated. We produce dictionaries according to $SF\in \{7,8,9,10\}$, and store them in the cloud server. 

\begin{figure}[hbt]
	\centering
    \vspace{-6pt}
	\begin{subfigure}{0.46\linewidth}
		\centering
		\includegraphics[width=\linewidth]{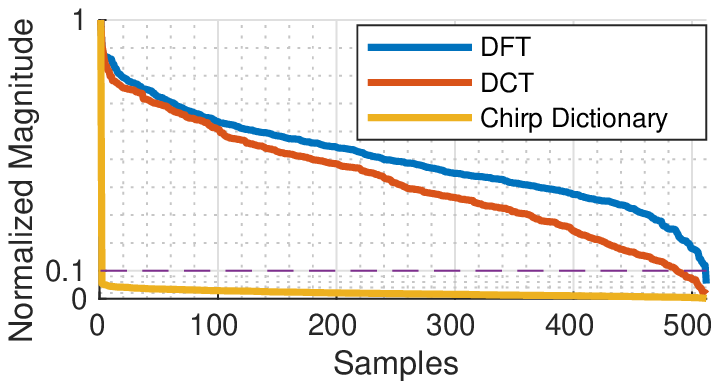}
		\caption{Synchronized}
		\label{fig:sparsity:sync}
	\end{subfigure}
	\hfill
	\begin{subfigure}{0.46\linewidth}
		\centering
		\includegraphics[width=\linewidth]{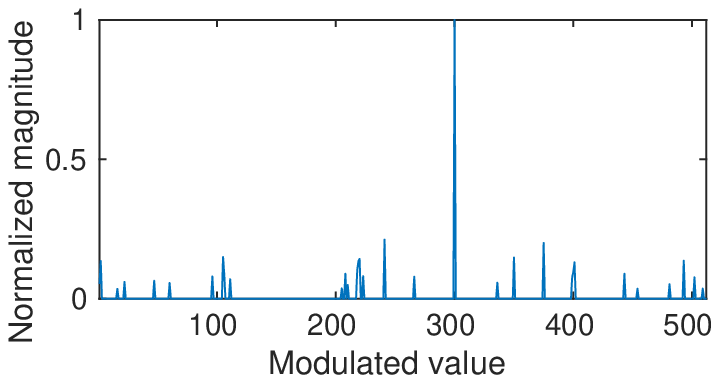}
		\caption{Solving with magnitude}
		\label{fig:solving:magnitude}
	\end{subfigure}
	\hfill
	\begin{subfigure}{0.46\linewidth}
		\centering
		\includegraphics[width=\linewidth]{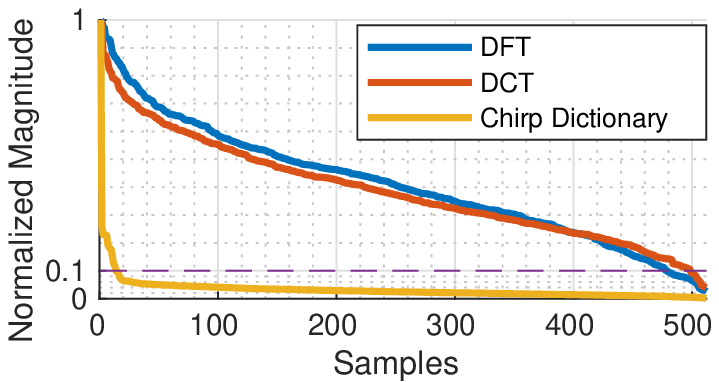}
		\caption{Unsynchronized}
		\label{fig:sparsity:unsync}
	\end{subfigure}
	\hfill
	\begin{subfigure}{0.46\linewidth}
		\centering
		\includegraphics[width=\linewidth]{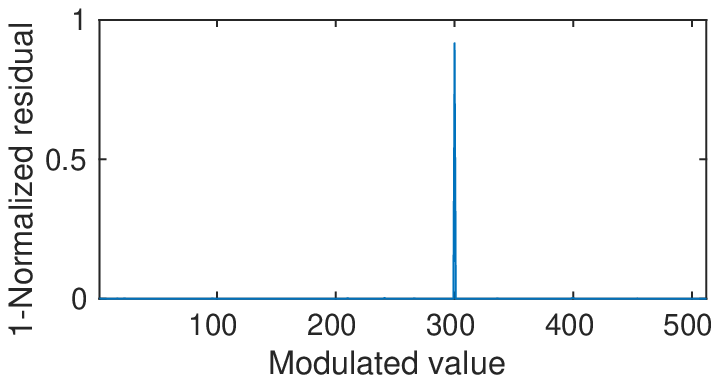}
		\caption{Solving with residual}
		\label{fig:solving:residual}
	\end{subfigure}
    \vspace{-5pt}
	\caption{(a) Sparsity for synchronized symbols based on DFT, DCT and the proposed chirp dictionary. It is more sparse in the proposed dictionary ($\Psi$) than the DFT and DCT by two orders-of-magnitude. The dashed line denotes the threshold for the coefficients with significant magnitude (0.1). (b) Sparse approximation with magnitude (Sec.~\ref{section:decoding}).  (c) Signal sparsity for unsynchronized chirps, less sparse than synchronized chirps but more sparse than the DFT and DCT. (d) Sparse approximation with residuals (Sec.~\ref{section:decoding}); the residual domain is more sparse than the magnitude domain. 
	} 
% 		\textcolor{red}{To bring back residual figure with phase, Fig.5 in NSDI}}
	\label{fig:sparsitylevel}
	\vspace{-15pt}
\end{figure}

As a comparison with DFT, DCT, and the proposed chirp dictionary, Fig.~\ref{fig:sparsity:sync} and~\ref{fig:sparsity:unsync} show the sparsity of typical synchronized and unsynchronized LoRa symbols ($SF=9$) with channel noise 
in different domains by sorting the samples by order of magnitude. The fastest decay characteristic (or the smallest $K$) is observed in the proposed
dictionary ($\Psi$), and therefore offers the most sparse representation; which means that the most accurate approximations (or LoRa symbol value estimations) can be obtained in this dictionary by using the smallest number of measurements $M$ (Eq.~(\ref{eq:M})). The sparsity in synchronized symbols is slightly better than the unsynchronized, which means that the accuracy in recovering synchronized symbols is better than the unsynchronized. The figure also shows that the proposed $\Psi$ has two-order-of-magnitude fewer significant coefficients
(e.g., the normalized magnitude is larger than 0.1) than those of DFT and DCT. 

For the down-chirps in PHY, % in SFD as shown in Fig.~\ref{fig:lora-phy}, 
similar dictionaries can be obtained by replacing  $\mathbf{U}$ with a matrix with a diagonal made of an ideal up-chirp. Due to the fact that most chirps in LoRa PHY are up-chirps, we first solve $\ell_1$-minimization with the up-chirp dictionary, and then try the down-chirp dictionary if no satisfactory result is obtained. Both dictionaries have similar features and performance. For brevity, we skip the discussion of the down-chirp dictionary.

% \begin{figure*}[hbt]
% 	\centering
% 	\begin{subfigure}{0.30\linewidth}
% 		\centering
% 		\includegraphics[width=\linewidth]{figures_new/2_dictionary_sparsity/s2_sparsity_sync_SF9.eps}
% 		\caption{Synchronized symbols}
% 		\label{fig:sparsity:sync}
% 	\end{subfigure}
% % 	\qquad
% 	\hfill
% 	\begin{subfigure}{0.30\linewidth}
% 		\centering
% 		\includegraphics[width=\linewidth]{figures_new/2_dictionary_sparsity/s2_sparsity_unsync_SF9.eps}
% 		\caption{Unsynchronized symbols}
% 		\label{fig:sparsity:unsync}
% 	\end{subfigure}
% 	\hfill
% 	\begin{subfigure}{0.34\linewidth}
% 		\centering
% 		\includegraphics[width=\linewidth]{figures_new/2_dictionary_sparsity/s2_solving_residual.eps}
% 		\caption{Solving with residual}
% 		\label{fig:solving:residual}
% 	\end{subfigure}

% 	\caption{(a) Signal sparsity based on DFT, DCT and proposed chirp dictionary. The signal is more sparse in the proposed dictionary ($\Psi$) than the DFT and DCT by two orders-of-magnitude. The dashed line denotes the threshold for the coefficients with significant magnitude (0.1). The signal is synchronized. (b) Signal sparsity for unsynchronized chirp, less sparse than synchronized chirp but more sparse than the DFT and DCT by two orders-of-magnitude. (c) Solving sparse approximation with residual (Sec.~\ref{section:decoding}), and the residual domain (lower figure) is more sparse than magnitude domain (upper figure). 
% 	} 
% % 		\textcolor{red}{To bring back residual figure with phase, Fig.5 in NSDI}}
% 	\label{fig:sparsitylevel}
% \end{figure*}

\subsubsection{Measurement Matrix}
\label{subsubsec:Phi}

%where $M$ and $N$ are the number of rows and columns of  measurement matrix $\Phi \in \mathbb{C}^{M \times N} $ respectively. 
As discussed in CS theory~\cite{Baraniuk2007, Donoho2006,Candes2006}, zero-mean Gaussian matrix and balance symmetric random Bernoulli matrix achieve favorable compression performance. 
For the computational efficiency on embedded devices, we choose random Bernoulli($\pm1$) as the measurement matrix $\Phi$ with a fixed seed that is shared by both gateways and the cloud server.

Each symbol has $N=2^{SF}$ samples, i.e. $N=$128, 256, 512, 1024 for $SF=7, 8, 9, 10$ respectively. If we process $SF$ separately, we have to compress PHY four times with $\Phi_{7}$, $\Phi_{8}$, $\Phi_{9}$, $\Phi_{10}$  for each $SF$, which is against our motivation for compression. 
To solve this problem, we only measure with $\Phi_{7}$. For $SF=8$, we can simply concatenate two compressed vectors from $\Phi_{7}$.
%which is equivalent to a larger matrix whose diagonal is composed of two $\Phi_{7}$. 
Similarly, we concatenate four compressed vectors for $SF=9$ and eight compressed vectors for $SF=10$. 

Thus, the gateway simply compresses every 128 samples with $\Phi_{7}$ for each channel, and in the cloud the server concatenates compressed vectors for solving different $SF$s.

%However, we follow the method proposed in~\cite{Xue2017,Rana2015} to achieve even better performance than conventional Gaussian or Bernoulli measurement matrix.
%The proposed method composes the measurement matrix $\Phi_S$ based on the Singular Value Decomposition (SVD) of the dictionary $\Psi$, and $\Phi_S$ is the transpose of the matrix that is made of the first $M$ largest singular vectors of $\Psi$.

\subsubsection{Compression ratio}
\label{sec:sub:compressionratio}

\begin{figure*}[htb]
	\centering
	\includegraphics[width=\textwidth]{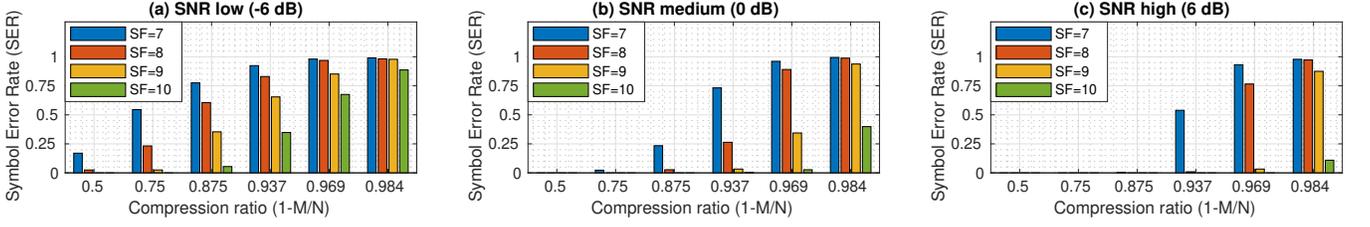}
	\caption{Simulation with \emph{synchronized} symbols: SER affected by compression ratio and SNR for different SF}
	\label{fig:simu:sync}
	\vspace{-5pt}
\end{figure*}

\begin{figure*}[htb]
	\centering
	\includegraphics[width=\textwidth]{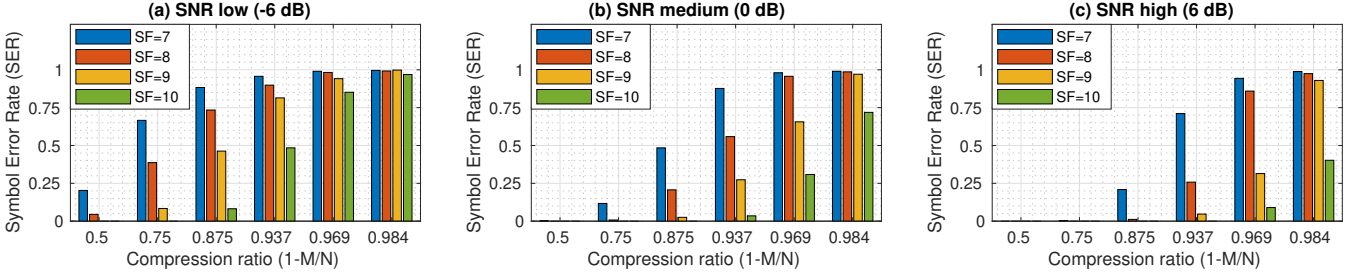}
	\caption{Simulation with \emph{unsynchronized} symbols: SER affected by compression ratio and SNR for different SF}
	\label{fig:simu:unsync}
	\vspace{-5pt}
\end{figure*}

Compression ratios are defined by Eq. (\ref{equ:alpha}), and thus a smaller $M$ results in a better compression ratio. Theoretically, $M$ should be bounded on its lower end by Eq. (\ref{eq:M}). However, the noise from the original signal is hidden in compressed vectors, which may make it challenging to recover the original signals (i.e.,  $\ell_1$ minimization 
algorithm fails 
to solve Eq. (\ref{eq:l1_new})) . 
%Davenport et. al. state that CS is sensitive to signal noise, exhibiting a 3dB SNR loss per octave of subsampling \cite{Davenport2012}.
Thus, $M$ is not only bounded by Eq. (\ref{eq:M}), but is also affected by the signal SNR. 
We perform a simulation to investigate this phenomenon. As $N=2^{SF}$ is an exponent of 2, to simplify the DSP process, $M$ is selected among exponents of 2 (e.g., 16, 32, 64, etc.). Here, we define low, medium and high SNRs as -6, 0 and 6 dB.

Fig.~\ref{fig:simu:sync} shows that higher $SF$s outperform their lower 
counterparts, and increasing SNR can improve the compression ratio. When SNR is high, $SF=9$ and $SF=7$  can be compressed to $1/16$ and $1/8$ respectively 
without significant Symbol Error Rates (SERs), and the compression ratio is mainly bounded by Eq. (\ref{eq:M}). When SNR is medium and low, $SF=9$ can be compressed to $1/16$ and $1/4$ respectively without significant SERs, and the compression ratio is mainly affected by SNR. 
\begin{table}[h]
    \vspace{-5pt}
	\caption{Reliable compression ratios based on simulations represented by $M/N$}
    \vspace{-5pt}
	\label{table:alpha1}	
	\begin{tabular}{lllll}
		\hline
		& SF7 & SF8  & SF9  & SF10 \\ \hline
		low SNR (-6 dB)    & 1   & 1/2    & 1/4  & 1/8  \\
		medium SNR (0 dB) & 1/4 & 1/8  & 1/16 & 1/32 \\
		high SNR (6 dB)  & 1/8 & 1/16 & 1/32 & 1/32 \\ \hline
	\end{tabular}
    \vspace{-3pt}
\end{table}

We summarize $M/N$ in Table~\ref{table:alpha1} to represent the acceptable compression ratio $\alpha$ if SER is small (e.g., $\leq0.04$). Then, the empirical compression ratio based on Fig.~\ref{fig:simu:sync} and Table~\ref{table:alpha1} can be derived as:
\vspace{-2pt}
\begin{equation}
\label{eq:empiricalalpha}
\alpha = max \{ min \{ 1-2^{-\floor{\frac{SNR_{dB}}{3}+SF-5}} ,1-\frac{2 \cdot SF}{2^{SF}} \}, 0 \}.
\end{equation}

For unsynchronized symbols, as shown in Fig.~\ref{fig:simu:unsync}, the performance is slightly poorer than that of the synchronized symbols. An unsynchronized symbol is composed of fractions of two consecutive chirp symbols (i.e. the last few samples from the first chirp and the first few samples from the second chirp).
%, see Fig.~\ref{fig:demod_22}
Thus, sparsity $K$ is increased from 1 to 2, and the lower bound Eq. (\ref{eq:M}) is slightly larger than that of the synchronized symbols. We modified Eq. (\ref{eq:empiricalalpha}) to select
an appropriate compression ratio for unsynchronized symbols accordingly:
\vspace{-2pt}
\begin{equation}
\label{eq:empiricalalpha2}
\alpha = max \{ min \{ 1-2^{-\floor{\frac{SNR_{dB}}{3}+SF-6}} ,1-\frac{4  (SF-1)}{2^{SF}} \}, 0 \}.
\end{equation}

\section{\emph{Nephelai} in the cloud}
\label{section:inthecloud}

\subsection{Decoding (Single Gateway)}
\label{section:decoding}

Most conventional $\ell_1$-minimization algorithms require real-valued vectors and dictionaries, while communication systems always use complex values for I/Q modulation.
% To this end, we firstly transform our problem 
To solve this problem we transform the vectors from complex-valued to real-valued as, 
\begin{align}
\mathbf{y'} &= [ \Re\{\mathbf{y}\}^T \quad \Im\{\mathbf{y}\}^T]^T \\
\mathbf{s'} &= [ \Re\{\mathbf{s}\}^T \quad \Im\{\mathbf{s}\}^T]^T \label{eq:complexs} \\ 
\Theta' &= \begin{bmatrix}
                \Re\{\Theta\} & -\Im\{\Theta\}\\
                \Im\{\Theta\} & \Re\{\Theta\}
            \end{bmatrix}
\end{align}

where $\Theta = \Phi \Psi$. Then, we solve the problem with a real-valued $\ell_1$-minimization algorithm for Eq. (\ref{eq:l1_new}) as,
\begin{equation} 
    \label{eq:l1_new2}
	\hat{\mathbf{s}}' = \arg\min \left\| \mathbf{s'} \right\|_1 \quad  s.t. \quad  \left\| \mathbf{y'} - \Theta' \mathbf{s'}  \right\|_2 < \epsilon
\end{equation}

After obtaining the sparse vector $ \hat{\mathbf{s}}'$ with Eq.~(\ref{eq:l1_new2}), we recover the complex-valued sparse vector $\mathbf{s_{opt}}$ by reversing Eq.~(\ref{eq:complexs}), and thus we solve not only the magnitude but the phase of the chirp symbol. 

Instead of using FFT for demodulation as described in Sec.~\ref{section:demodulation}, we proceed to estimate the most likely value $\lambda$ by using residual $r$.
%$\mathbf{r}$.
The residual for symbol candidate $i \in$ \{0, 1, ..., $2^{SF}$ -1\} 
%with phase offset $j \in \{0, \delta\!\varphi, ... ,(R-1)\delta\!\varphi \} $
%$j \in \{0, 1,..., \Delta\}$ 
is:
\begin{equation} 
\label{eq:residual}
    \vspace{-1pt}
    r^{(i)} (\mathbf{y}) =   \left\| \mathbf{y} - \Phi \Psi \delta^{(i)}(\mathbf{s_{opt}})  \right\|_2, \forall i
\end{equation}

where operator $\delta^{(i)}:\mathbb{R}^{D}\rightarrow\mathbb{R}^{D}$ indicates a vector containing the only coefficient related to candidates $i$ (the coefficients related to other candidates are set to be zeros). Then the final symbol estimation is determined by:
\vspace{-2pt}
\begin{equation}
\label{eq:solve_lambda}
    \hat{\lambda} =  \operatorname*{argmin}_{i} r^{(i)} (\mathbf{y}) ,  \forall i
\end{equation}
\vspace{-2pt}
i.e., the $\lambda$ with the minimal residual representing the modulation value.
Fig.~\ref{fig:solving:residual} shows the
result of \emph{Nephelai} decoding with Eq.~(\ref{eq:residual}) for a noisy chirp symbol. The highest peak (i.e., $1 - r^{(i)}$, suppose $r^{(i)}$ is normalized) represents the modulated value (e.g., 300) of the LoRa symbol correctly. Note that in $\mathbf{s_{opt}}$, the phase of the highest peak may be used for radio-based ranging, which is beyond the scope of this paper.

\subsection{Joint Decoding}
\label{section:jointdecoding}

We have discussed how \emph{Nephelai} recovers value $\lambda$ from
single compressed measurement $\mathbf{y}$. In this section, we discuss
how \emph{Nephelai} exploits spatial diversity for gateways and improves performance with joint decoding. 

\begin{figure}[htb]
	\centering
	\includegraphics[width=0.64\linewidth]{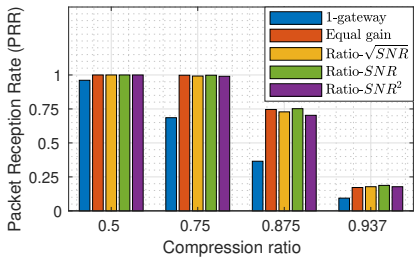}
	\caption{Joint decoding algorithm comparison}
	\label{fig:revision3:joint}
    \vspace{-7pt}
\end{figure}

% ============= revision ==============
Suppose that we have $G$ gateways, and each gateway captures a transmitted copy of the same LoRa symbol independently. Next, \emph{Nephelai} estimates the SNR level $\gamma_g$ and produces residuals $r_{g}^{(i)}$ ( $g \in \{0, 1, ... G-1\}$) for $G$ gateways with Eq.~(\ref{eq:residual}). 
One of the ways to fuse these residuals among gateways is to perform a weighted summation. Based on the selection of combining weights, we have four algorithms: 1) weighted equally, aka. equal gain combining (EGC); 2) weighted by the $\sqrt{SNR}$; 3) weighted by the $SNR$ aka. the maximum ratio combining (MRC), and 4) weighted by the $SNR^2$. We evaluate the algorithms with collected samples by four gateways (further discussion in Sec. \ref{subsubsec:jointdecoding}), and the results are shown in Fig.~\ref{fig:revision3:joint}. All algorithms succeed in improving the PRR, and the algorithm weighted by the $SNR$ has the best performance especially when the compression ratio is high. Thus, we choose the MRC algorithm with SNR $\gamma_g$ as the weight in the following evaluation.
% ============= revision ==============

Following this, the final symbol estimation is determined by:
\vspace{-3pt}
\begin{equation}
\label{eq:residual_min}
\hat{\lambda} = \arg\min_{i} \sum_{g = 0}^{G-1}{\gamma_g r_{g}^{(i)} (\mathbf{y})} ,\forall i
  %  \lambda =  \arg\min_i \sum_{p = 0}^{P-1}{\chi_i^p (\mathbf{y})}, \forall i,
\end{equation}

% We also try other fusion techniques such as taking the absolute summation of the minimum residual of a LoRa symbol (i.e., min $(r_{j,g}^{(i)}$), $\forall i, \forall g$, which represents
% LoRa symbol $i$ from the most dominate radio path between the transmitter and gateway $g$, but the results show insignificant difference 
% (see Fig.~\ref{fig:compare_min_and_add_residual} later). 
% The algorithm of \emph{Nephelai} joint decoding 
Nephelai's joint decoding algorithm can be found in Algorithm~\ref{algo:jointdecoding}. 
\vspace{-5pt}

\begin{algorithm}
	\DontPrintSemicolon % Some LaTeX compilers require you to use \dontprintsemicolon instead 
	\KwIn{$M$-length measurements $\{\mathbf{y}_g\}_{g=0..G-1}$, estimated SNR $\{\gamma_g\}_{g=0..G-1}$ }
	\KwOut{An integer $\lambda$, the decoding result}

	\For{$g \gets 0$ \textbf{to} $G-1$}{
		$\mathbf{s}_g \gets $ solve $\ell_1$ minimization($\mathbf{y}_g$, $\Theta$, $\epsilon$)\;
		\For{$i \gets 0$ \textbf{to} $2^{SF}-1$}{
		$ r_{g}^{(i)} (\mathbf{y}) =   \left\| \mathbf{y_g} - \Theta \delta^{(i)}(\mathbf{s_{g}})  \right\|_2 $ \\
        }
	}
	$\lambda \gets  \argmin_{i} \sum_{g=0}^{G-1} \gamma_g r_{g}^{(i)} $

	\Return{$\lambda$}
	\caption{{\sc joint-decoding}}
	\label{algo:jointdecoding}
\end{algorithm}

\vspace{-8pt}
\section{Prototype Implementation}
\label{sec:prototype}

\begin{figure}[hbt]
	\centering
	\includegraphics[width=0.71\linewidth]{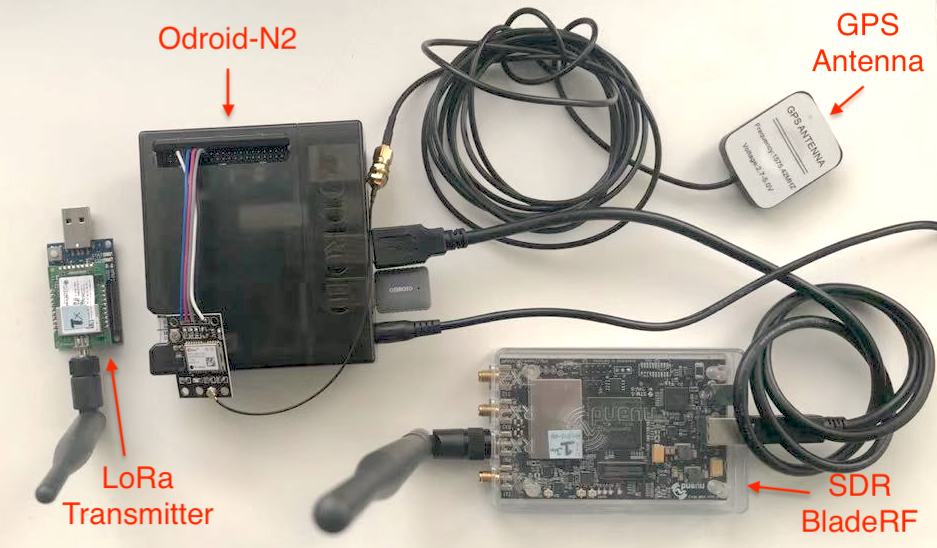}
	\vspace{-7pt}
    \caption{\emph{Nephelai} gateway and a LoRa transmitter}
	\label{fig:clora-hardware}
    \vspace{-10pt}
\end{figure}

\textbf{The Edge Gateway} The \emph{Nephelai} gateway shown in Fig.~\ref{fig:clora-hardware} has a radio front-end to capture signal samples on given LoRa channels, and an embedded computer to pre-process and compress the received signal samples before uploading to the cloud. In our prototype, we select BladeRF 2.0 SDR as the radio front-end to capture radio signals on LoRaWAN uplink channels (e.g., 902 MHz to 915 MHz in the USA). The output of SDR is a stream of $I$ and $Q$ components, which can be regarded as complex values where $I$ denotes real and $Q$ denotes imaginary parts respectively. 
The SDR can sample up to 61.44 mega samples per second (MSps), which are capable of capturing all the information in the whole 13 MHz upstream spectrum for USA defined by LoRaWAN. The Nyquist sampling rate for one channel is 125 kHz for complex samples (i.e. 250 kHz for real samples), and therefore the sample rate for 64 channels is 8 MSps (note the 75 kHz guard band between consecutive 125kHz channels, meaning that 8 MHz is for LoRa channels on a 13 MHz spectrum). 
%We define a pair of I and Q components as one radio sample. 

The SDR is connected to a Odroid-N2 (6-core single board computer with quad-core Cortex-A73@1.8GHz and dual-core Cortex-A53@1.9GHz) via a USB 3.0 port, through which the LoRa radio samples are transferred. Next, the Odroid-N2 processes (see Sec.~\ref{section:architecture})
and compresses (see Sec.~\ref{subsec:symbolcompression}) the samples before transferring them to the cloud server. The sampling rate of our prototype is 13 MHz, which is sufficient to cover the 13 MHz LoRaWAN spectrum. 
Without loss of generality, we demonstrate the compression performance of \emph{Nephelai} in a single LoRa uplink channel. If one single channel is compressible,
%so does other 63 channels.  
so are 63 other channels.

We design and implement the software for \emph{Nephelai} gateways, called \emph{gr-Nephelai} based on the open-source software-defined ratio platform GNU-Radio. The frequency conversion and low pass filter shown in Fig.~\ref{fig:clora-architecture} are implemented in C++ and complied with single instruction multiple data (SIMD) optimization. Although there are 64 parallel branches in Fig.~\ref{fig:clora-architecture}, we implement one block for all 64 channels instead of one block for each of the 64 channels to reduce the handover between blocks. 
%The taps of the low pass filter is selected as 
The low-pass filter taps are selected as 47 to maintain real-time performance. The passband is designed to be 275 kHz, which works well to avoid inter-channel interference. When the gateway is running at full capacity (processing 64 channels), the overall CPU usage is approximately 60\%.

\textbf{The transmitter} We program Multitech mDot\footnote{
MDot datasheet. %Hyperlink is removed according to the submission instruction.
\url{https://www.multitech.com/brands/multiconnect-mdot}
}, 
which comprises a LoRa wireless chip (SX1272), to periodically transmit 4 predefined bytes. 
%In physical layer, 4 bytes are modulated to 16 chirps for payload. 
The mDot with STM32F411RET uses 31 mA @100 MHz in the maximum power setting.

\textbf{The Cloud Server} Although the \emph{Nephelai} cloud server can be any kind of general server, we use a 12-core CPU, 32 GB RAM and Nvidia 2070 GPU server in our 
prototype. It can perform $\ell_1$-minimization algorithms for joint sparse LoRa signal reconstruction (i.e., LoRa packet decoding, see Sec.~\ref{section:jointdecoding}) in real-time.

\section{Evaluation}

\label{section:evaluation}

\subsection{Goals, Metrics and Methodologies}
\label{subsec:goals}

\begin{figure} [htb]
	\vspace{-8pt}
	\centering
	\includegraphics[width=0.99\linewidth]{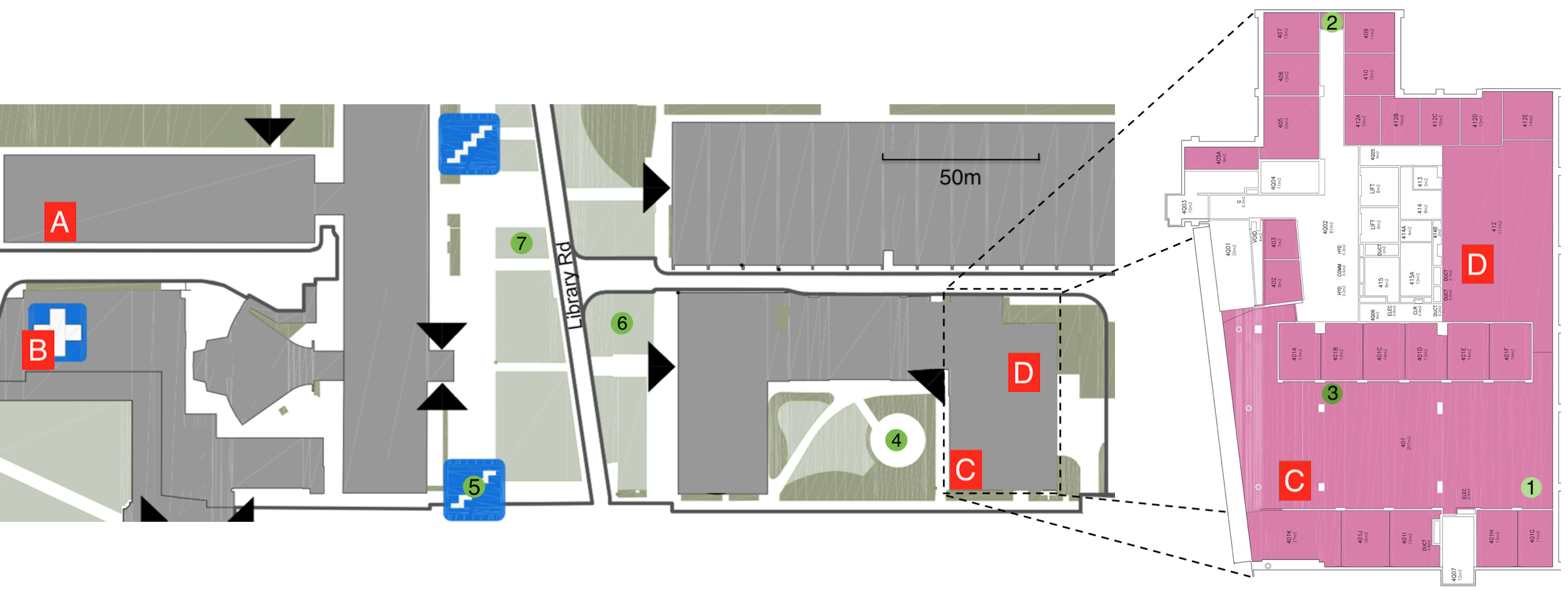}
	\vspace{-8pt}
	\caption{\emph{Nephelai} test-bed on our campus. The gateways are marked with the letter A/B/C/D and stationed inside buildings near windows to simulate a customer-deployed scenario. The transmitters (motes) are labeled from 1 to 7, and marked with green circles. Mote-1 is on the same floor (the 4th floor) as gateway C; mote-2 is on the 3rd floor; mote-3 is hidden in the basement, 5 floors below gateway C. Motes-4/5/6/7 are placed outdoor without any cover.}
	\label{fig:floorplan}
	\vspace{-8pt}
\end{figure}

We deployed a \emph{Nephelai} testbed with four \emph{Nephelai} gateways (see Sec.~\ref{sec:prototype}) on our campus as shown in Fig.~\ref{fig:floorplan}, where gateways are connected to a \emph{Nephelai} cloud server (see Sec.~\ref{sec:prototype}) via Wi-Fi. We programmed seven mDots (see Fig.~\ref{fig:clora-hardware}) as LoRa motes to periodically transmit predefined LoRa packets with power from 2 dBm to 14 dBm. We installed the LoRa motes in several representative positions in the campus to emulate real applications, and collected LoRa radio samples with each gateway simultaneously. During our evaluation, we collected more than one million LoRa chirp symbols among SF7 to SF10 to evaluate the performance of \emph{Nephelai}. 

We deployed LoRa motes to emulate real use cases. Mote-1 was an indoor temperature and humidity sensor; mote-2 acted as a passive infrared sensor (PIR), which functioned as an occupancy detector for the warehouse; mote-3 behaved as a smart water meter; mote-4 represented a simple outdoor weather station; mote-5 was attached
% to the handrail of steps represents a people counter; 
to a stair handrail and counted people; and mote-6 and mote-7 measured the soil's humidity to control a watering system for the lawn. 
%However, in this evaluation, we are not 
In this evaluation we were not interested in application data 
% but only 
but instead focused on PHY compression and potential battery lifetime improvement with joint decoding.

\emph{Nephelai} is designed to implement the physical layer compression for cloud-assisted LoRa demodulation/decoding and to potentially improve transmitters' energy efficiency. Therefore, the \textbf{goals} of our evaluation were:

\begin{enumerate}
    \item to study whether \emph{Nephelai} can reduce the network bandwidth of the front-haul in LPWAN C-RAN,
    
    \item to study the impact of compression ratio ($\alpha$) on the system's performance, and
        
    \item to study whether \emph{Nephelai} can demonstrate similar energy improvements for the LoRa transmitter as the state-of-the-art LPWAN C-RAN, but with fewer front-haul data rates.
\end{enumerate}

The \textbf{metric} for network bandwidth reduction is bits per second (bps), and that for energy reduction is battery lifetime extension. For \textbf{methodologies}, firstly, on the symbol level we evaluate how SNR and compression ratios affect SER in order to compare these with the simulation in Sec.~\ref{sec:sub:compressionratio}. And then on packet level, we evaluated the PRR for single gateway scenarios with three LoRa motes and different power transmission levels. Furthermore, we evaluated the joint processing gain with four gateways and four transmitters to demonstrate that an equivalent SNR improvement can be achieved as the state-of-the-art \cite{Dongare2018}, i.e. to extend the battery lifetime to approximately 1.7x (equivalent to 2.3 dB SNR improvement) with four gateways, but with greater PHY compression. As there are different SFs resulting in different PRRs, we assumed that each SF was equal likely to be selected, and we calculated the expected PRR by averaging the PRRs of all SFs.

%%%%%%%%%%%%%%%%%%%%%%%%%%%%%%%%%%%%%%%%%%%%%%%%%%%%%%%%%%%%%%%%%%%%%%%%%%%%%%%%%%%%%%%%%%%%%%%%%%%%%%
%%%%%%%%%%%%%%%%%%%%%%%%%%%%%%%%%%%%%%%%%%%%%%%%%%%%%%%%%%%%%%%%%%%%%%%%%%%%%%%%%%%%%%%%%%%%%%%%%%%%%%
\vspace{-5pt}
\subsection{Empirical Results}
\label{subsec:emperical}

%%%%%%%%%%%%%%%%%%%%%%%%%%%%%%%%%%%%%%%%%%%%%%%%%%%%%%%%%%%%%%%%%%%%%%%%%%%%%%%%%%%%%%%%%%%%%%%%%%%%%%

\subsubsection{Compression ratio}
\label{sec:eva:compressionratio}

\begin{figure*}[htb]
	\centering
	\includegraphics[width=\textwidth]{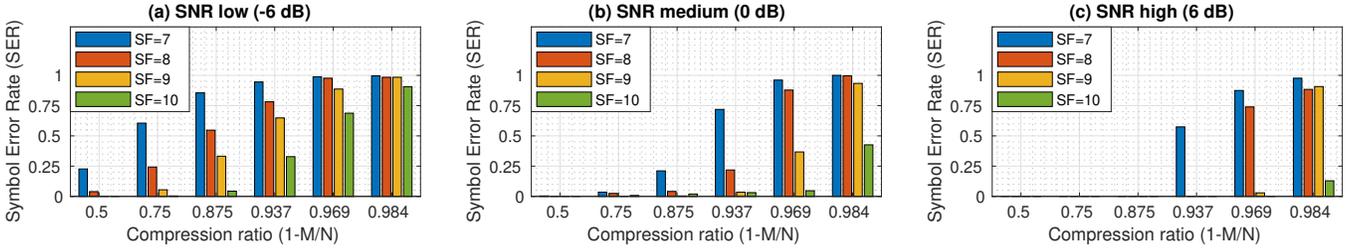}
	\caption{Synchronized symbols from testbed: SER affected by compression ratio and SNR for different SFs}
	\vspace{-5pt}
	\label{fig:eva:sync}
\end{figure*}

\begin{figure*}[htb]
	\centering
	\includegraphics[width=\textwidth]{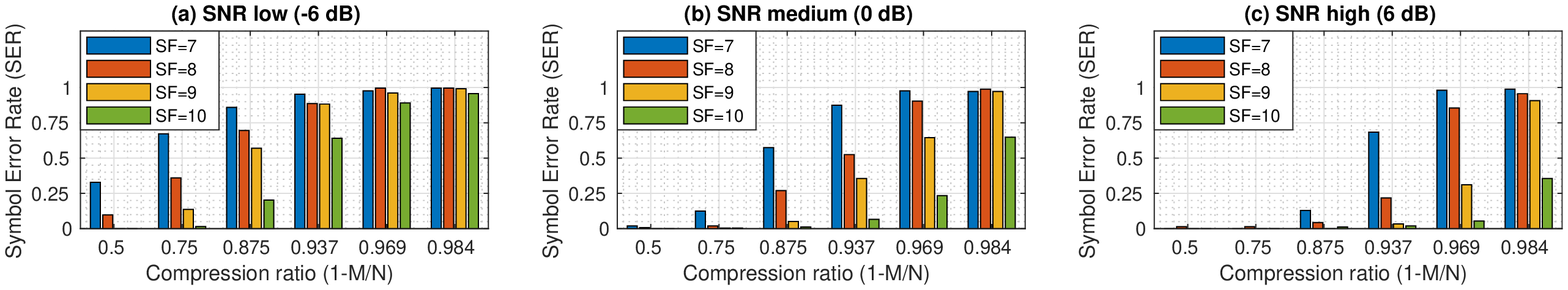}
	\caption{Unynchronized symbols from testbed: SER affected by compression ratio and SNR for different SFs}
	\label{fig:eva:unsync}
\end{figure*}

As discussed in Sec.~\ref{subsubsec:Phi}, the compression ratio ($\alpha$) is calculated using the dimension of measurement matrix $\Phi \in \mathbb{C}^{M \times N}$ (see Eq.~(\ref{eq:y})). 
In this section, we are only interested in how SNR affected the compression ratios, and in evaluating the compression ratio determination equations (i.e., Eq. (\ref{eq:empiricalalpha}) and (\ref{eq:empiricalalpha2})) for synchronized and unsynchronized symbols. 
We programmed motes-1/2/3 to transmit with power varying from 2 dB to 14 dB, and collected  50,000 synchronized and unsynchronized symbols respectively. 
We grouped symbols with respect to their low (-6 dB), medium (0 dB) and high (6 dB) SNR.
Fig.~\ref{fig:eva:sync} and \ref{fig:eva:unsync} compare the compression performance of different SFs and SNRs based on the symmetric Bernoulli matrix($\Phi$) of $\pm 1$ and our proposed chirp dictionary $\Psi$ (see Sec.~\ref{subsec:symbolcompression}). 
For example, for medium SNR (0 dB, i.e. the signal energy is equivalent to the noise floor) with synchronized symbols in Fig.~\ref{fig:eva:sync}, SF9 achieves an SER below 0.04 with a compression ratio of 93.7\%. This represents approximately 16 times the bandwidth reduction in the C-RAN front-haul. 

With a small SER value (e.g., $\leq 0.04$) as the reliable transmission threshold, we can summarize that the evaluation matches the simulation, when referring to $M/N$ in Table~\ref{table:alpha2} based on Fig.~\ref{fig:eva:sync}, which compares Table~\ref{table:alpha2} to Table~\ref{table:alpha1}.
Therefore, we can use Eq.~(\ref{eq:empiricalalpha}) in compression ratio selection. We observed similar patterns in
the results of unsynchronized symbols in Eq.~(\ref{eq:empiricalalpha2}), however we omit the discussion here for brevity.

\begin{figure*}[htb]
	\centering
	\includegraphics[width=0.95\textwidth]{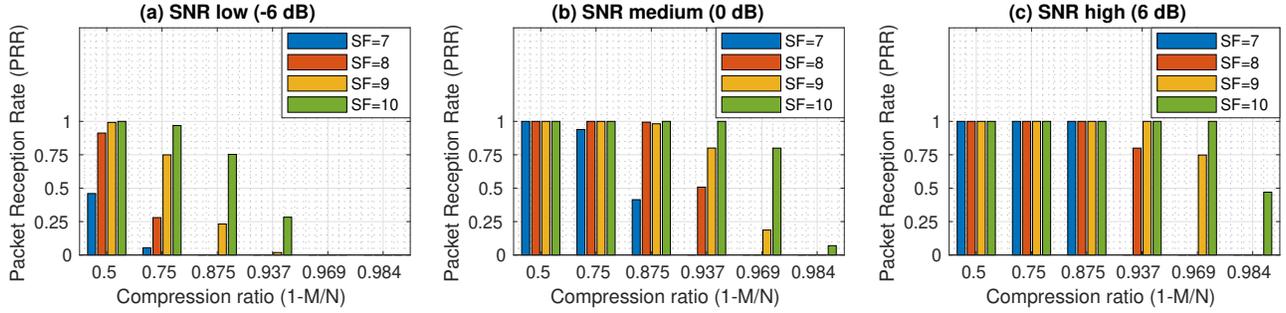}
	\caption{PRR affected by SNR, SFs and compression ratios for synchronized symbols/packets}
	\label{fig:revision2:prr}
\end{figure*}

% ============ revision ===============
% “Poor SER as a cost of the compression/demodulation (fig. 10/11). Can you report more results with regard the PRR (of typical payload length), especially for the low SNR range, so it will clearly show your trade-off between compression ratio and LoRa coverage; otherwise you should clearly highlight limitations with regard those aspects in the paper;”
Furthermore, we performed PRR evaluation for synchronized packets with different SNRs, SFs and compression ratios as shown in Fig.~\ref{fig:revision2:prr}. The LoRa packets transmitted in the evaluation had fixed length and their payloads consisted of 4 bytes (equivalent to 8 symbols). 
We defined PRR 75\% as the threshold for reliable transmission~\cite{le2007design} and used it in our compression ratio selection.
With the PRR criteria, Fig.~\ref{fig:revision2:prr} implies a similar compression ratio selection as that with SER in Table~\ref{table:alpha2}. Thus, we can use Eq.~(\ref{eq:empiricalalpha}) in compression ratio selection. For unsynchronized symbols, similar to the discussion with SER, Eq.~(\ref{eq:empiricalalpha2}) is used for compression ratio selection.

% ============ revision ===============

\begin{table}[]
	\caption{Reliable compression ratio based on testbed collected data represented by $M/N$.} %The difference to the simulation results in Table~\ref{table:alpha1} is highlighted in bold.
	\vspace{-3pt}
	\label{table:alpha2}
	\begin{tabular}{lllll}
		\hline
		& SF7 & SF8  & SF9  & SF10 \\ \hline
		low SNR (-6 dB)   & 1   & 1/2    & 1/4  & 1/8  \\
		medium SNR (0 dB) & 1/4 & 1/8  & 1/16 & 1/32 \\
		high SNR (6 dB)  & 1/8 & 1/16 & 1/32 & 1/32 \\ \hline
	\end{tabular}
	\vspace{-13pt}
\end{table}

In summary, compared to the benchmark of lossless algorithm LZ77 that achieves a compression ratio of 7.5\% (see Sec.~\ref{subsec:gzip} for more details), the proposed approach can improve the compression ratio by approximately 10 times, depending on the parameter settings. For example, when SNR is high, a compression ratio up to 93.7\% can be achieved for most SFs. Therefore, \emph{Nephelai} achieves a significant reduction in traffic between gateways and the cloud server, which makes the cloud-assisted LoRa decoding scheme more scalable.

\vspace{-6pt}

\subsubsection{The performance of single gateway}

\label{subsubsec:singledecoding}
% Evaluation for single gateway

\begin{figure*}[htb]
	\centering
	\includegraphics[width=0.98\textwidth]{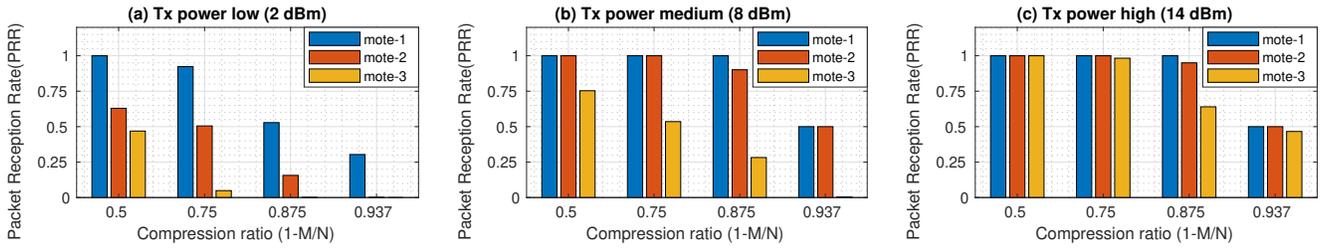}
% 	\caption{Single gateway evaluation with 3 transmitters for all spreading factors shows PRR is affected by compression ratios in different power transmission scenarios for \textbf{synchronized} symbols. Transmitter mote-1 was placed on the same floor as the gateway; mote-2 was placed one floor (steel and concrete building) under the gateway; mote-3 was placed in the basement, 5 floors under the gateway (see Fig.\ref{fig:floorplan} for transmitter deployment detail).}
	\caption{The single gateway evaluation with 3 transmitters and \textbf{synchronized} symbols shows that PRR is affected by compression ratios in different power transmission scenarios. Motes-1/2/3 were placed according to Fig.\ref{fig:floorplan}.}
	\label{fig:eva:single:sync}
\end{figure*}
	
% \begin{figure*}[htb]
% 	\centering
% 	\includegraphics[width=0.95\textwidth]{figures_new/5_evasingle/s5_unsync_lab_SF_combined.eps}
% 	\caption{Single gateway evaluation with identical settings as Fig.~\ref{fig:eva:single:sync} but for \textbf{unsynchronized} symbols (see  Fig.~\ref{fig:floorplan} for transmitter deployment detail).}
% 	\label{fig:eva:single:unsync}
% \end{figure*}

In the single gateway evaluation using a real case, our goal was to compress PHY without PRR degradation. As discussed in Sec.~\ref{section:compression}, over-compression means that the $\ell_1$ minimization 
algorithm fails 
to solve Eq. (\ref{eq:l1_new2}), which increases SERs and decreases PRRs.
%CS causes a 3 dB SNR loss per octave of subsampling. However, in the campus where we deploy the testbed, SNR is relatively high, while scalability is the main concern, and the key solution for scalability is to compress PHY as discussed in Sec.~\ref{section:introduction}. 

Firstly, as shown in Fig.~\ref{fig:floorplan}, LoRaWAN transmitter motes-1, 2 and 3 were installed in a fixed position and were programmed to transmit 4 bytes with different spreading factors ($SF=7,8,9,10$) at 2 dBm, 8 dBm and 14 dBm respectively. We collect packets via one gateway in either synchronized or unsynchronized mode. 
Secondly, with the algorithm proposed in Sec.~\ref{section:decoding}, we calculated the PRR for different compression ratios. 
Instead of SER, we were more interested in PRR which describes the performance of end-to-end data transmissions. For example, if PRR is halved, the energy required to successfully deliver one packet is doubled, as the embedded node needs to transmit the packet twice. Therefore, the battery lifetime is halved. It is evident that PRR is more intuitive than SER in describing battery lifetime. 
%We average PRR for all SFs to get an expected PRR as discussed in Sec.~\ref{subsec:goals}.

% Fig.~\ref{fig:eva:single:sync} shows that synchronized symbols can improve compression by approximately one octave when compared to the unsynchronized symbols as shown in \ref{fig:eva:single:unsync}. 
In our synchronized scenario, the compression ratio of 87.5\% for motes-1 and 2 produced more than 90\% PRR when power transmission was medium.  For mote-3 in the basement, the compression ratio of 75\% produced more than 50\% PRR. We note that mote-3 was over-compressed with the compression ratio of 87.5\% because the PRR is only 30\% (see Fig.~\ref{fig:eva:single:sync}). Increasing power transmission could have increased the compression ratio for mote-3 from 75\% to 87.5\%, allowing it to maintain its PRR above 50\% (Fig.~\ref{fig:eva:single:sync}(c)). According to the mDot datasheet,
% \footnote{Multitech mDot datasheet: % \url{https://www.multitech.com/documents/publications/data-sheets/86002171.pdf}}
increasing power transmission from medium to high consumes 3.7\% extra energy, which provides another acceptable option for scalability improvement. 

In summary, Fig.~\ref{fig:eva:single:sync} 
% and~\ref{fig:eva:single:unsync} 
shows that PRR does not decrease with appropriate compression ratios, and increasing power transmission can improve the compression performance of \emph{Nephelai}. 
Therefore, if all motes transmit at 14 dBm, we can select 87.5\% as the compression ratio. For 64 channels, only $64 \times 24 \times 125000 \times (1-0.875) = 24$ Mbps is required for a single gateway in LPWAN C-RAN. Consequently, such a gateway can operate with bandwidth-limited Internet connections, widely extending the deployment region and application scenarios.
% \vspace{-0.3cm}

%%%%%%%%%%%%%%%%%%%%%%%%%%%%%%%%%%%%%%%%%%%%%%%%%%%%%%%%%%%%%%%%%%%%%%%%%%%%%%%%%%%%%%%%%%%%%%%%%%%%%%

\subsubsection{The performance of joint decoding}
\label{subsubsec:jointdecoding}

% \textcolor{red}{rewrite}

\begin{figure*}[hbt]
	\centering
	\includegraphics[width=0.95\textwidth]{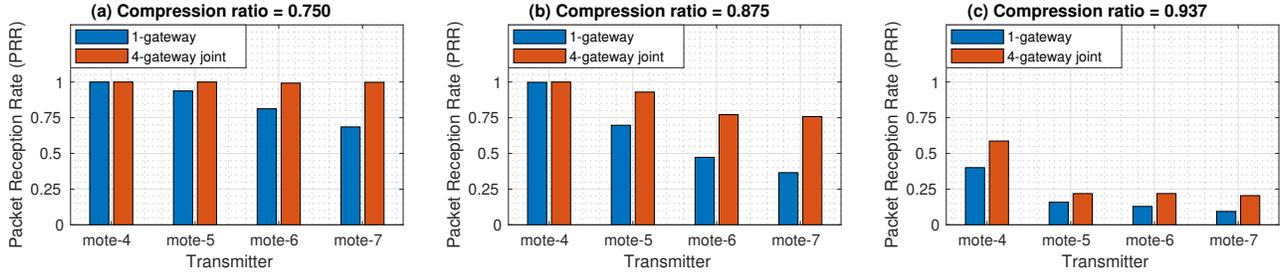}
	\caption{PRR improvement by 4-gateway joint decoding with compression ratio}
	%\textcolor{red}{replace the figure with combined PRR for SF=7,8,9,10, as the previous section, and draw conclusion based on new figure afterwards}
	\label{fig:eva_prr_improvement}
\end{figure*}

Compressing PHY without PRR degradation is possible as shown in Sec.~\ref{subsubsec:singledecoding} above. In this section, we evaluate the improvement of PRR with joint decoding under compression. Our goal was to achieve an equivalent performance to the state-of-the-art Charm system (i.e. 2.3 dB SNR improves with four gateways, see  Sec.~\ref{section:introduction} for the details), but with less front-haul bandwidth between the gateway and the cloud. 

Firstly, we programmed motes-4,5,6 and 7 to be in synchronized mode and to send 4 byte messages periodically with high transmission power\footnote{We define 14 dBm as high transmission power in this paper, but in fact 14 dBm is a moderate choice compared to the maximum 22 dBm.} (14 dBm). We collected LoRa radio samples simultaneously via gateways-A,B,C and D with different compression ratios (see Sec.~\ref{subsec:goals} and Fig.~\ref{fig:floorplan} for testbed deployment in details). The number of packets for each SF was equal.
Secondly, we calculated the PRR for single gateway decoding and coherent joint decoding with 4 gateways (according to the algorithm discussed in Sec.~\ref{section:inthecloud} under different compression ratios). 
We averaged PRR for all SFs to get an expected PRR as discussed previously in Sec.~\ref{subsec:goals}.

Fig.~\ref{fig:eva_prr_improvement} shows how much improvement can be seen by joint decoding with four gateways compared to a single gateway. For battery-powered LoRa motes, the expected energy consumption per packet is reversely proportional to the PRR, and thus the expected battery lifetime is proportional to the PRR.
When the compression ratio was 87.5\%, mote-4 had PRR above 99\% (since the position of mote-4 was very close to one of the gateways), while motes-5,6 and 7 had poor PRR with a single gateway. After joint decoding with four gateways, the PRR of mote-5 was improved from 70\% up to 93\%, while mote-6 went from 47\% to 77\%, and mote-7 went from 36\% to 76\%. The improvement factors are 1.33, 1.64 and 2.11 respectively, and the average is about 1.70. Therefore, on average, joint decoding extends battery lifetime to approximately 1.70 with four gateways when the compression ratio is 87.5\%. 
We note that
% the minimal improved PRR is when 
the least improved PRR occurred when the compression ratio is 75\%, which is equivalent to a good quality, low power
wireless link with a high C-RAN bit rate. Therefore, 87.5\% is the recommended trade-off between compression ratio and PRR.

For compression up to 93.7\%, single gateways experience severe packet loss for each mote. After joint decoding, the PRR of mote-4 was improved from 40\% to 58\%, while mote-5 improved from 16\% to 22\%, mote-6 from 13\% to 22\%, and mote-7 from 10\% to 20\%.
%The improvement is 1.38, 1.38, 1.64 and 1.9 times, which means equivalent battery life improvement.
However, this compression ratio is not recommended because most of the PRRs are still poor (i.e., less than 50\%) even with joint decoding. 
Particularly, increasing the compression ratio from 87.5\% to 93.7\% for mote-4 causes PRR to decrease from more than 99\% to 40\%, 
meant that the mote had a shorter battery lifetime by approximately 60\%.
Finally, we note that when the compression ratio is 75\%, with joint decoding, all PRRs are increased to more than 99\%.

In summary, \emph{Nephelai} with 4 gateways improves the PPR and the battery lifetime of a LoRa transmitter by 1.7 times on average, with the recommended compression ratio of 87.5\%  compared to a single gateway, which is equivalent to 2.3 dB SNR improvement ($ 10 log_{10} 1.7$). The compression ratio of 87.5\% also means that the PHY is compressed from 3 Mbps down to 375 kbps for one channel, while that of Charm is 2.25 Mbps per channel (see Sec.~\ref{section:introduction} for details). %It shows 
This demonstrates that \emph{Nephelai} has similar functionality in improving the battery lifetime of embedded IoT devices as Charm~\cite{Dongare2018}, while \emph{Nephelai} reduces the bandwidth between gateways and the cloud by $1-0.375/2.25=83.3\%$.

\subsubsection{Cloud computing overhead}
Solving $\ell_1$ minimization is computationally intensive, but can be handled with parallel implementation using multi-threading, GPU, FPGA, etc. in the cloud. If the demodulation of one symbol is performed in real-time, and the delay caused by data transmission and computation (from the gateway to the cloud, and back to the gateway) meets the LoRaWAN requirement for an ACK, the \emph{Nephelai} system is feasible.

We evaluated cloud computing overhead by performing single-threading tests with MATLAB on Intel Core i7-8700 CPU @ 3.20GHz with 32GB RAM for 1000 times calculation per case as shown in Table.~\ref{table:revision1:overhead}. The worst case is SF10 with a 50\% compression ratio. One symbol for SF10 can be solved in less than 500 ms with a single thread. The length of one symbol for SF10 is 8.2 ms, and in 500 ms the gateway can receive at most 61 of these symbols. 
%Thus, one may use a 64-core server in the cloud and 
Therefore, a 64 core server can be used in the cloud to dispatch demodulation tasks to each core in order to obtain real-time demodulation within 500 ms. For other SFs and compression ratios, the computational demand is even lower. Note that the computation can be further optimised for higher efficiency.

LoRaWAN has a relatively loose requirement for ACK delays due to low bit rates (e.g., 300 bps). There is a parameter called ACK\_TIMEOUT in the LoRaWAN settings with a default value of "$2 \pm 1$s ( i.e., a random delay between 1 and 3 seconds)”. The demodulation latency is less than 500ms as discussed above, and the Internet latency is typically less than 100 milliseconds one way. Processing latency caused by gateways and radio propagation delays can be ignored. Thus, an ACK can easily be generated in one second to meet the LoRaWAN requirements discussed above.

\begin{table}[h]
	\vspace{-4pt}
	\caption{$\ell_1$-minimization overhead testing for different SFs and compression ratios. Unit: millisecond.} 
	\vspace{-4pt}
	\label{table:revision1:overhead}
	\begin{tabular}{lllll}
		\hline
		$\alpha$ & SF7 & SF8  & SF9  & SF10 \\ 
		\hline
0.5   & 5.1$\pm$ 2.5 &	10.8$\pm$2.7 &	66.8  $\pm$61.9 &	297.7$\pm$180.5 \\
0.75  & 2.0$\pm$	0.6 &	4.7$\pm$	1.0 &	16.3$\pm$13.7 &	71.2$\pm$32.6 \\
0.875 & 1.0$\pm$	0.2 &	2.2$\pm$	0.4 &	5.5$\pm$ 3.3  &	16.1$\pm$6.1 \\
0.937 & 0.6$\pm$	0.1 &	1.1$\pm$	0.2 &	2.3$\pm$ 0.9  &	5.1$\pm$1.8 \\
		\hline
	\end{tabular}
	\vspace{-6pt}
\end{table}

\subsubsection{Influence of concurrent transmission}

Theoretically, multi-channel concurrent transmission may reduce the system's performance by leaking energy as noise to other channels. However, through our evaluation, we have found that concurrent transmission does not cause system degradation. 

We established a LoRa transmitter that sent packets with SF=8 and a packet length of 41.5 ms every 50 ms periodically in one 125kHz channel, and another transmitter that sent in the neighbouring channels. We calculated the PRR based on the collected samples in different interference environments: no interference, concurrent transmission on a +200kHz channel, concurrent transmission on a +400kHz channel, ... , and concurrent transmission on a +1000kHz channel. 
%Fig.~\ref{fig:revision4:adjacent} 
Our evaluation results show that no significant PRR reduction is caused by concurrent transmissions.
If we have a well designed filter for each 125kHz channel, the noise caused by concurrent transmissions can be prevented. In summary, \emph{Nephelai} is robust against the interference caused by concurrent transmissions.

% \begin{figure}[htb]
% 	\centering
% 	\includegraphics[width=0.99\linewidth]{figures_revision/re4_PRR_adjacent.eps}
% 	\caption{PRR affected by concurrent transmissions.}
% 	\label{fig:revision4:adjacent}
% \end{figure}

% \input{contents/80-LimitationFuturework.tex}

\section{Conclusion}
\label{section:conclusion}
We introduce \emph{Nephelai}, which is based on CS-theory, to reduce the
bandwidth requirement between edge gateways and the cloud server for cloud-assisted
LoRaWAN. \emph{Nephelai} exploits: 1) the physical layer structure of LoRa symbols for a custom designed \emph{dictionary} to
significantly improve its compression performance, 
2) the relationship between compression ratios, SNR and SFs to select an appropriate compression ratio, and
3) radio signal spatial diversity by joint decoding to improve the PRR as well as the battery lifetime for end devices. 
Our empirical results with an edge gateway prototype consisting 
of SDR and Odroid-N2 show that \emph{Nephelai} can reduce 
traffic between gateways and cloud servers by up to 93.7\% and can significantly improve
the scalability of cloud assisted LoRaWAN. 
%Furthermore, 
%by combining radio signal received in 4 gateway via joint decoding, \emph{Nephelai}
%can double the battery lifetime of embedded LoRaWAN transmitters for data
%rates DR0 to DR3.
%We design and implement a C-RAN for jointly decoding LoRa packet, called \emph{Nephelai}, with Cs technique to reduce network bandwidth. Our evaluation show a strong scalability is achieved. Only 1.5 Mbps is required for one gateway (DR0 to DR3), and the battery life of transmitters can be doubled.

% \begin{acks}
% \end{acks}

\bibliographystyle{ACM-Reference-Format}
\bibliography{mendeley_export.bib}


\begin{thebibliography}{43}

%%% the Web address.

\ifx \showCODEN    \undefined \def \showCODEN     #1{\unskip}     \fi
\ifx \showDOI      \undefined \def \showDOI       #1{#1}\fi
\ifx \showISBNx    \undefined \def \showISBNx     #1{\unskip}     \fi
\ifx \showISBNxiii \undefined \def \showISBNxiii  #1{\unskip}     \fi
\ifx \showISSN     \undefined \def \showISSN      #1{\unskip}     \fi
\ifx \showLCCN     \undefined \def \showLCCN      #1{\unskip}     \fi
\ifx \shownote     \undefined \def \shownote      #1{#1}          \fi
\ifx \showarticletitle \undefined \def \showarticletitle #1{#1}   \fi
\ifx \showURL      \undefined \def \showURL       {\relax}        \fi
% invisible to TeX
\providecommand\bibfield[2]{#2}
\providecommand\bibinfo[2]{#2}
\providecommand\natexlab[1]{#1}
\providecommand\showeprint[2][]{arXiv:#2}

\bibitem[\protect\citeauthoryear{Adelantado, Vilajosana, Tuset-Peiro, Martinez,
  Melia-Segui, and Watteyne}{Adelantado et~al\mbox{.}}{2017}]%
        {Adelantado2017}
\bibfield{author}{\bibinfo{person}{Ferran Adelantado}, \bibinfo{person}{Xavier
  Vilajosana}, \bibinfo{person}{Pere Tuset-Peiro}, \bibinfo{person}{Borja
  Martinez}, \bibinfo{person}{Joan Melia-Segui}, {and} \bibinfo{person}{Thomas
  Watteyne}.} \bibinfo{year}{2017}\natexlab{}.
\newblock \showarticletitle{{Understanding the Limits of LoRaWAN}}.
\newblock \bibinfo{journal}{\emph{IEEE Communications Magazine}}
  \bibinfo{volume}{55}, \bibinfo{number}{9} (\bibinfo{year}{2017}),
  \bibinfo{pages}{34--40}.
\newblock
\showISBNx{978-0-9949886-0-7}
\showISSN{01636804}
\urldef\tempurl%
\url{https://doi.org/10.1109/MCOM.2017.1600613}
\showDOI{\tempurl}
\showeprint[arxiv]{1607.08011}


\bibitem[\protect\citeauthoryear{Baraniuk}{Baraniuk}{2007}]%
        {Baraniuk2007}
\bibfield{author}{\bibinfo{person}{Richard Baraniuk}.}
  \bibinfo{year}{2007}\natexlab{}.
\newblock \showarticletitle{{Compressive Sensing [Lecture Notes]}}.
\newblock \bibinfo{journal}{\emph{IEEE Signal Processing Magazine}}
  \bibinfo{volume}{24}, \bibinfo{number}{4} (\bibinfo{date}{jul}
  \bibinfo{year}{2007}), \bibinfo{pages}{118--121}.
\newblock
\showISBNx{1053-5888 VO - 24}
\showISSN{1053-5888}
\urldef\tempurl%
\url{https://doi.org/10.1109/MSP.2007.4286571}
\showDOI{\tempurl}


\bibitem[\protect\citeauthoryear{Beyene, Jantti, Tirkkonen, Ruttik, Iraji,
  Larmo, Tirronen, and Torsner}{Beyene et~al\mbox{.}}{2017}]%
        {Beyene2017}
\bibfield{author}{\bibinfo{person}{Yihenew~Dagne Beyene}, \bibinfo{person}{Riku
  Jantti}, \bibinfo{person}{Olav Tirkkonen}, \bibinfo{person}{Kalle Ruttik},
  \bibinfo{person}{Sassan Iraji}, \bibinfo{person}{Anna Larmo},
  \bibinfo{person}{Tuomas Tirronen}, {and} \bibinfo{person}{Johan Torsner}.}
  \bibinfo{year}{2017}\natexlab{}.
\newblock \showarticletitle{{NB-IoT Technology Overview and Experience from
  Cloud-RAN Implementation}}.
\newblock \bibinfo{journal}{\emph{IEEE Wireless Communications}}
  \bibinfo{volume}{24}, \bibinfo{number}{3} (\bibinfo{date}{jun}
  \bibinfo{year}{2017}), \bibinfo{pages}{26--32}.
\newblock
\showISBNx{1536-1284}
\showISSN{15361284}
\urldef\tempurl%
\url{https://doi.org/10.1109/MWC.2017.1600418}
\showDOI{\tempurl}


\bibitem[\protect\citeauthoryear{Bor, Roedig, Voigt, and Alonso}{Bor
  et~al\mbox{.}}{2016}]%
        {Bor}
\bibfield{author}{\bibinfo{person}{Martin Bor}, \bibinfo{person}{Utz Roedig},
  \bibinfo{person}{Thiemo Voigt}, {and} \bibinfo{person}{Juan~M Alonso}.}
  \bibinfo{year}{2016}\natexlab{}.
\newblock \showarticletitle{{Do LoRa low-power wide-area networks scale?}}. In
  \bibinfo{booktitle}{\emph{MSWiM 2016 - Proceedings of the 19th ACM
  International Conference on Modeling, Analysis and Simulation of Wireless and
  Mobile Systems}}. \bibinfo{pages}{59--67}.
\newblock
\showISBNx{9781450345026}
\urldef\tempurl%
\url{https://doi.org/10.1145/2988287.2989163}
\showDOI{\tempurl}


\bibitem[\protect\citeauthoryear{Brennan and D.G.BRENNAN}{Brennan and
  D.G.BRENNAN}{1959}]%
        {Brennan1959}
\bibfield{author}{\bibinfo{person}{D.~G. Brennan} {and}
  \bibinfo{person}{D.G.BRENNAN}.} \bibinfo{year}{1959}\natexlab{}.
\newblock \showarticletitle{{Linear Diversity Combining Technique}}.
\newblock \bibinfo{journal}{\emph{Proceedings of the IRE}}
  \bibinfo{volume}{10}, \bibinfo{number}{6} (\bibinfo{date}{jun}
  \bibinfo{year}{1959}), \bibinfo{pages}{1075--1102}.
\newblock
\showISBNx{0018-9219}
\showISSN{0096-8390}
\urldef\tempurl%
\url{https://doi.org/10.1109/JRPROC.1959.287136}
\showDOI{\tempurl}


\bibitem[\protect\citeauthoryear{Cand{\`e}s, Romberg, and Tao}{Cand{\`e}s
  et~al\mbox{.}}{2006}]%
        {Candes2006}
\bibfield{author}{\bibinfo{person}{Emmanuel~J. Cand{\`e}s},
  \bibinfo{person}{Justin Romberg}, {and} \bibinfo{person}{Terence Tao}.}
  \bibinfo{year}{2006}\natexlab{}.
\newblock \showarticletitle{{Robust uncertainty principles: Exact signal
  reconstruction from highly incomplete frequency information}}.
\newblock \bibinfo{journal}{\emph{IEEE Transactions on Information Theory}}
  \bibinfo{volume}{52}, \bibinfo{number}{2} (\bibinfo{date}{sep}
  \bibinfo{year}{2006}), \bibinfo{pages}{489--509}.
\newblock
\showISBNx{0018-9448 VO - 52}
\showISSN{00189448}
\urldef\tempurl%
\url{https://doi.org/10.1109/TIT.2005.862083}
\showDOI{\tempurl}
\showeprint[arxiv]{math/0409186}


\bibitem[\protect\citeauthoryear{Centenaro, Vangelista, Zanella, and
  Zorzi}{Centenaro et~al\mbox{.}}{2016}]%
        {centenaro2016long}
\bibfield{author}{\bibinfo{person}{Marco Centenaro}, \bibinfo{person}{Lorenzo
  Vangelista}, \bibinfo{person}{Andrea Zanella}, {and} \bibinfo{person}{Michele
  Zorzi}.} \bibinfo{year}{2016}\natexlab{}.
\newblock \showarticletitle{Long-range communications in unlicensed bands: The
  rising stars in the IoT and smart city scenarios}.
\newblock \bibinfo{journal}{\emph{IEEE Wireless Communications}}
  \bibinfo{volume}{23}, \bibinfo{number}{5} (\bibinfo{year}{2016}),
  \bibinfo{pages}{60--67}.
\newblock


\bibitem[\protect\citeauthoryear{Checko, Christiansen, Yan, Scolari, Kardaras,
  Berger, and Dittmann}{Checko et~al\mbox{.}}{2015}]%
        {Checko2015}
\bibfield{author}{\bibinfo{person}{Aleksandra Checko},
  \bibinfo{person}{Henrik~L. Christiansen}, \bibinfo{person}{Ying Yan},
  \bibinfo{person}{Lara Scolari}, \bibinfo{person}{Georgios Kardaras},
  \bibinfo{person}{Michael~S. Berger}, {and} \bibinfo{person}{Lars Dittmann}.}
  \bibinfo{year}{2015}\natexlab{}.
\newblock \showarticletitle{{Cloud RAN for Mobile Networks - A Technology
  Overview}}.
\newblock \bibinfo{journal}{\emph{IEEE Communications Surveys {\&} Tutorials}}
  \bibinfo{volume}{17}, \bibinfo{number}{1} (\bibinfo{year}{2015}),
  \bibinfo{pages}{405--426}.
\newblock
\showISBNx{1553-877X VO - PP}
\showISSN{1553-877X}
\urldef\tempurl%
\url{https://doi.org/10.1109/COMST.2014.2355255}
\showDOI{\tempurl}


\bibitem[\protect\citeauthoryear{{China Mobile}}{{China Mobile}}{2011}]%
        {ChinaMobile2011}
\bibfield{author}{\bibinfo{person}{{China Mobile}}.}
  \bibinfo{year}{2011}\natexlab{}.
\newblock \showarticletitle{{C-RAN: the road towards green RAN}}.
\newblock \bibinfo{journal}{\emph{White Paper, ver 2.5}}  \bibinfo{volume}{5}
  (\bibinfo{year}{2011}), \bibinfo{pages}{15--16}.
\newblock


\bibitem[\protect\citeauthoryear{Dongare, Narayanan, Gadre, Luong, Balanuta,
  Kumar, Iannucci, and Rowe}{Dongare et~al\mbox{.}}{2018}]%
        {Dongare2018}
\bibfield{author}{\bibinfo{person}{Adwait Dongare}, \bibinfo{person}{Revathy
  Narayanan}, \bibinfo{person}{Akshay Gadre}, \bibinfo{person}{Anh Luong},
  \bibinfo{person}{Artur Balanuta}, \bibinfo{person}{Swarun Kumar},
  \bibinfo{person}{Bob Iannucci}, {and} \bibinfo{person}{Anthony Rowe}.}
  \bibinfo{year}{2018}\natexlab{}.
\newblock \showarticletitle{{Charm: Exploiting Geographical Diversity Through
  Coherent Combining in Low-power Wide-area Networks}}. In
  \bibinfo{booktitle}{\emph{Proceedings of the 17th ACM/IEEE International
  Conference on Information Processing in Sensor Networks}}
  \emph{(\bibinfo{series}{IPSN '18})}. \bibinfo{publisher}{IEEE Press},
  \bibinfo{address}{Piscataway, NJ, USA}, \bibinfo{pages}{60--71}.
\newblock
\showISBNx{978-1-5386-5298-5}
\urldef\tempurl%
\url{https://doi.org/10.1109/IPSN.2018.00013}
\showDOI{\tempurl}


\bibitem[\protect\citeauthoryear{Donoho}{Donoho}{2006}]%
        {Donoho2006}
\bibfield{author}{\bibinfo{person}{David~L. Donoho}.}
  \bibinfo{year}{2006}\natexlab{}.
\newblock \showarticletitle{{For most large underdetermined systems of linear
  equations the minimal l1-norm solution is also the sparsest solution}}.
\newblock \bibinfo{journal}{\emph{Communications on Pure and Applied
  Mathematics}} \bibinfo{volume}{59}, \bibinfo{number}{6} (\bibinfo{date}{jun}
  \bibinfo{year}{2006}), \bibinfo{pages}{797--829}.
\newblock
\showISSN{0010-3640}
\urldef\tempurl%
\url{https://doi.org/10.1002/cpa.20132}
\showDOI{\tempurl}


\bibitem[\protect\citeauthoryear{Farrell}{Farrell}{2018}]%
        {Farrell2018}
\bibfield{author}{\bibinfo{person}{S Farrell}.}
  \bibinfo{year}{2018}\natexlab{}.
\newblock \bibinfo{booktitle}{\emph{{Low-Power Wide Area Network (LPWAN)
  Overview}}}.
\newblock \bibinfo{type}{{T}echnical {R}eport}. \bibinfo{institution}{RFC
  Editor}. \bibinfo{pages}{1--43} pages.
\newblock
\showISSN{2070-1721}
\urldef\tempurl%
\url{https://doi.org/10.17487/RFC8376}
\showDOI{\tempurl}


\bibitem[\protect\citeauthoryear{Gadre, Narayanan, Luong, Rowe, Iannucci, and
  Kumar}{Gadre et~al\mbox{.}}{2020}]%
        {gadre2020frequency}
\bibfield{author}{\bibinfo{person}{Akshay Gadre}, \bibinfo{person}{Revathy
  Narayanan}, \bibinfo{person}{Anh Luong}, \bibinfo{person}{Anthony Rowe},
  \bibinfo{person}{Bob Iannucci}, {and} \bibinfo{person}{Swarun Kumar}.}
  \bibinfo{year}{2020}\natexlab{}.
\newblock \showarticletitle{Frequency Configuration for Low-Power Wide-Area
  Networks in a Heartbeat}. In \bibinfo{booktitle}{\emph{17th $\{$USENIX$\}$
  Symposium on Networked Systems Design and Implementation ($\{$NSDI$\}$ 20)}}.
  \bibinfo{pages}{339--352}.
\newblock


\bibitem[\protect\citeauthoryear{Georgiou and Raza}{Georgiou and Raza}{2017}]%
        {Georgiou2017}
\bibfield{author}{\bibinfo{person}{Orestis Georgiou} {and}
  \bibinfo{person}{Usman Raza}.} \bibinfo{year}{2017}\natexlab{}.
\newblock \showarticletitle{{Low Power Wide Area Network Analysis: Can LoRa
  Scale?}}
\newblock \bibinfo{journal}{\emph{IEEE Wireless Communications Letters}}
  \bibinfo{volume}{6}, \bibinfo{number}{2} (\bibinfo{date}{oct}
  \bibinfo{year}{2017}), \bibinfo{pages}{162--165}.
\newblock
\showISSN{21622345}
\urldef\tempurl%
\url{https://doi.org/10.1109/LWC.2016.2647247}
\showDOI{\tempurl}
\showeprint[arxiv]{1610.04793}


\bibitem[\protect\citeauthoryear{Ghanaatian, Afisiadis, Cotting, and
  Burg}{Ghanaatian et~al\mbox{.}}{2019}]%
        {ghanaatian2019lora}
\bibfield{author}{\bibinfo{person}{Reza Ghanaatian}, \bibinfo{person}{Orion
  Afisiadis}, \bibinfo{person}{Matthieu Cotting}, {and}
  \bibinfo{person}{Andreas Burg}.} \bibinfo{year}{2019}\natexlab{}.
\newblock \showarticletitle{LoRa digital receiver analysis and implementation}.
  In \bibinfo{booktitle}{\emph{ICASSP 2019-2019 IEEE International Conference
  on Acoustics, Speech and Signal Processing (ICASSP)}}. IEEE,
  \bibinfo{pages}{1498--1502}.
\newblock


\bibitem[\protect\citeauthoryear{Hoeller, Souza, L{\'o}pez, Alves,
  de~Noronha~Neto, and Brante}{Hoeller et~al\mbox{.}}{2018}]%
        {hoeller2018analysis}
\bibfield{author}{\bibinfo{person}{Arliones Hoeller},
  \bibinfo{person}{Richard~Demo Souza}, \bibinfo{person}{Onel L~Alcaraz
  L{\'o}pez}, \bibinfo{person}{Hirley Alves}, \bibinfo{person}{Mario de
  Noronha~Neto}, {and} \bibinfo{person}{Glauber Brante}.}
  \bibinfo{year}{2018}\natexlab{}.
\newblock \showarticletitle{Analysis and performance optimization of LoRa
  networks with time and antenna diversity}.
\newblock \bibinfo{journal}{\emph{IEEE Access}}  \bibinfo{volume}{6}
  (\bibinfo{year}{2018}), \bibinfo{pages}{32820--32829}.
\newblock


\bibitem[\protect\citeauthoryear{Knight}{Knight}{2016}]%
        {Knight2016}
\bibfield{author}{\bibinfo{person}{Matthew Knight}.}
  \bibinfo{year}{2016}\natexlab{}.
\newblock \showarticletitle{{Decoding LoRa : Realizing a Modern LPWAN with
  SDR}}. In \bibinfo{booktitle}{\emph{Proceedings of the 6th GNU Radio
  Conference,}}, Vol.~\bibinfo{volume}{1}.
\newblock


\bibitem[\protect\citeauthoryear{Le~Dinh, Hu, Sikka, Corke, Overs, and
  Brosnan}{Le~Dinh et~al\mbox{.}}{2007}]%
        {le2007design}
\bibfield{author}{\bibinfo{person}{Tuan Le~Dinh}, \bibinfo{person}{Wen Hu},
  \bibinfo{person}{Pavan Sikka}, \bibinfo{person}{Peter Corke},
  \bibinfo{person}{Leslie Overs}, {and} \bibinfo{person}{Stephen Brosnan}.}
  \bibinfo{year}{2007}\natexlab{}.
\newblock \showarticletitle{Design and deployment of a remote robust sensor
  network: Experiences from an outdoor water quality monitoring network}. In
  \bibinfo{booktitle}{\emph{32nd IEEE Conference on Local Computer Networks
  (LCN 2007)}}. IEEE, \bibinfo{pages}{799--806}.
\newblock


\bibitem[\protect\citeauthoryear{Liando, Gamage, Tengourtius, and Li}{Liando
  et~al\mbox{.}}{2019}]%
        {Liando2019}
\bibfield{author}{\bibinfo{person}{Jansen~C. Liando}, \bibinfo{person}{Amalinda
  Gamage}, \bibinfo{person}{Agustinus~W. Tengourtius}, {and}
  \bibinfo{person}{Mo Li}.} \bibinfo{year}{2019}\natexlab{}.
\newblock \showarticletitle{{Known and unknown facts of LoRa: Experiences from
  a large-scale measurement study}}.
\newblock \bibinfo{journal}{\emph{ACM Transactions on Sensor Networks}}
  \bibinfo{volume}{15}, \bibinfo{number}{2} (\bibinfo{year}{2019}).
\newblock
\showISSN{15504867}
\urldef\tempurl%
\url{https://doi.org/10.1145/3293534}
\showDOI{\tempurl}


\bibitem[\protect\citeauthoryear{Logan}{Logan}{1965}]%
        {logan1965properties}
\bibfield{author}{\bibinfo{person}{B. Logan}.} \bibinfo{year}{1965}\natexlab{}.
\newblock \emph{\bibinfo{title}{Properties of high-pass signals}}.
\newblock \bibinfo{thesistype}{Ph.D. Dissertation}. \bibinfo{school}{Columbia
  University}.
\newblock


\bibitem[\protect\citeauthoryear{{LoRa Alliance}}{{LoRa Alliance}}{2017}]%
        {LoRaAlliance2017}
\bibfield{author}{\bibinfo{person}{{LoRa Alliance}}.}
  \bibinfo{year}{2017}\natexlab{}.
\newblock \showarticletitle{{LoRaWAN 1.1 Specification}}.
\newblock \bibinfo{journal}{\emph{LoRa Alliance}} \bibinfo{number}{1.1}
  (\bibinfo{year}{2017}), \bibinfo{pages}{101}.
\newblock
\urldef\tempurl%
\url{https://lora-alliance.org/resource-hub/lorawantm-specification-v11}
\showURL{%
\tempurl}


\bibitem[\protect\citeauthoryear{Misra, Hu, Jin, Liu, {De Paula}, Wirstr{\"o}m,
  and Voigt}{Misra et~al\mbox{.}}{2014}]%
        {Misra2014}
\bibfield{author}{\bibinfo{person}{Prasant Misra}, \bibinfo{person}{Wen Hu},
  \bibinfo{person}{Yuzhe Jin}, \bibinfo{person}{Jie Liu},
  \bibinfo{person}{Amanda~Souza {De Paula}}, \bibinfo{person}{Niklas
  Wirstr{\"o}m}, {and} \bibinfo{person}{Thiemo Voigt}.}
  \bibinfo{year}{2014}\natexlab{}.
\newblock \showarticletitle{{Energy efficient GPS acquisition with
  Sparse-GPS}}. In \bibinfo{booktitle}{\emph{IPSN 2014 - Proceedings of the
  13th International Symposium on Information Processing in Sensor Networks
  (Part of CPS Week)}}. \bibinfo{publisher}{IEEE}, \bibinfo{pages}{155--166}.
\newblock
\showISBNx{9781479931460}
\urldef\tempurl%
\url{https://doi.org/10.1109/IPSN.2014.6846749}
\showDOI{\tempurl}


\bibitem[\protect\citeauthoryear{Misra, Ostry, Kottege, and Jha}{Misra
  et~al\mbox{.}}{2011}]%
        {Misra2011}
\bibfield{author}{\bibinfo{person}{Prasant~Kumar Misra},
  \bibinfo{person}{Diethelm Ostry}, \bibinfo{person}{Navinda Kottege}, {and}
  \bibinfo{person}{Sanjay Jha}.} \bibinfo{year}{2011}\natexlab{}.
\newblock \showarticletitle{{TWEET: an envelope detection based broadband
  ultrasonic ranging system}}. In \bibinfo{booktitle}{\emph{Proceedings of the
  14th ACM international conference on Modeling, analysis and simulation of
  wireless and mobile systems}}. \bibinfo{publisher}{ACM},
  \bibinfo{pages}{409--416}.
\newblock
\showISBNx{978-1-4503-0898-4}
\urldef\tempurl%
\url{https://doi.org/10.1145/2068897.2068967}
\showDOI{\tempurl}


\bibitem[\protect\citeauthoryear{Park, Simeone, Sahin, and {Shamai Shitz}}{Park
  et~al\mbox{.}}{2014}]%
        {Park2014}
\bibfield{author}{\bibinfo{person}{Seok~Hwan Park}, \bibinfo{person}{Osvaldo
  Simeone}, \bibinfo{person}{Onur Sahin}, {and} \bibinfo{person}{Shlomo {Shamai
  Shitz}}.} \bibinfo{year}{2014}\natexlab{}.
\newblock \showarticletitle{{Fronthaul compression for cloud radio access
  networks: Signal processing advances inspired by network information
  theory}}.
\newblock \bibinfo{journal}{\emph{IEEE Signal Processing Magazine}}
  \bibinfo{volume}{31}, \bibinfo{number}{6} (\bibinfo{date}{nov}
  \bibinfo{year}{2014}), \bibinfo{pages}{69--79}.
\newblock
\showISSN{10535888}
\urldef\tempurl%
\url{https://doi.org/10.1109/MSP.2014.2330031}
\showDOI{\tempurl}


\bibitem[\protect\citeauthoryear{Peng, Wang, Lau, and Poor}{Peng
  et~al\mbox{.}}{2015}]%
        {Peng2015}
\bibfield{author}{\bibinfo{person}{Mugen Peng}, \bibinfo{person}{Chonggang
  Wang}, \bibinfo{person}{Vincent Lau}, {and} \bibinfo{person}{H.~Vincent
  Poor}.} \bibinfo{year}{2015}\natexlab{}.
\newblock \showarticletitle{{Fronthaul-Constrained Cloud Radio Access Networks:
  Insights and Challenges}}.
\newblock \bibinfo{journal}{\emph{IEEE Wireless Communications}}
  \bibinfo{volume}{22}, \bibinfo{number}{2} (\bibinfo{date}{mar}
  \bibinfo{year}{2015}), \bibinfo{pages}{152--160}.
\newblock
\showISSN{15361284}
\urldef\tempurl%
\url{https://doi.org/10.1109/MWC.2015.7096298}
\showDOI{\tempurl}
\showeprint[arxiv]{1503.01187}


\bibitem[\protect\citeauthoryear{Polonelli, Brunelli, Marzocchi, and
  Benini}{Polonelli et~al\mbox{.}}{2019}]%
        {Polonelli2019}
\bibfield{author}{\bibinfo{person}{Tommaso Polonelli}, \bibinfo{person}{Davide
  Brunelli}, \bibinfo{person}{Achille Marzocchi}, {and} \bibinfo{person}{Luca
  Benini}.} \bibinfo{year}{2019}\natexlab{}.
\newblock \showarticletitle{{Slotted ALOHA on LoRaWAN-design, analysis, and
  deployment}}.
\newblock \bibinfo{journal}{\emph{Sensors (Switzerland)}} \bibinfo{volume}{19},
  \bibinfo{number}{4} (\bibinfo{year}{2019}).
\newblock
\showISSN{14248220}
\urldef\tempurl%
\url{https://doi.org/10.3390/s19040838}
\showDOI{\tempurl}


\bibitem[\protect\citeauthoryear{Rao and Lau}{Rao and Lau}{2015}]%
        {Rao2015}
\bibfield{author}{\bibinfo{person}{Xiongbin Rao} {and}
  \bibinfo{person}{Vincent~K.N. Lau}.} \bibinfo{year}{2015}\natexlab{}.
\newblock \showarticletitle{{Distributed fronthaul compression and joint signal
  recovery in cloud-RAN}}.
\newblock \bibinfo{journal}{\emph{IEEE Transactions on Signal Processing}}
  \bibinfo{volume}{63}, \bibinfo{number}{4} (\bibinfo{date}{feb}
  \bibinfo{year}{2015}), \bibinfo{pages}{1056--1065}.
\newblock
\showISSN{1053587X}
\urldef\tempurl%
\url{https://doi.org/10.1109/TSP.2014.2386290}
\showDOI{\tempurl}


\bibitem[\protect\citeauthoryear{Robyns, Peter, Wim, and William}{Robyns
  et~al\mbox{.}}{2017}]%
        {Robyns2017}
\bibfield{author}{\bibinfo{person}{Pieter Robyns}, \bibinfo{person}{Quax
  Peter}, \bibinfo{person}{Lamotte Wim}, {and} \bibinfo{person}{Thenaers
  William}.} \bibinfo{year}{2017}\natexlab{}.
\newblock \bibinfo{title}{{gr-lora: An efficient LoRa decoder for GNU Radio.}}
\newblock
\newblock
\urldef\tempurl%
\url{https://doi.org/10.5281/zenodo.853201}
\showDOI{\tempurl}


\bibitem[\protect\citeauthoryear{Robyns, Quax, Lamotte, and Thenaers}{Robyns
  et~al\mbox{.}}{2018}]%
        {robyns2018multi}
\bibfield{author}{\bibinfo{person}{Pieter Robyns}, \bibinfo{person}{Peter
  Quax}, \bibinfo{person}{Wim Lamotte}, {and} \bibinfo{person}{William
  Thenaers}.} \bibinfo{year}{2018}\natexlab{}.
\newblock \showarticletitle{A multi-channel software decoder for the LoRa
  modulation scheme}.
\newblock


\bibitem[\protect\citeauthoryear{Saari, bin Baharudin, Sillberg, Hyrynsalmi,
  and Yan}{Saari et~al\mbox{.}}{2018}]%
        {Saari2018}
\bibfield{author}{\bibinfo{person}{M. Saari}, \bibinfo{person}{A.~Muzaffar bin
  Baharudin}, \bibinfo{person}{P. Sillberg}, \bibinfo{person}{S. Hyrynsalmi},
  {and} \bibinfo{person}{W. Yan}.} \bibinfo{year}{2018}\natexlab{}.
\newblock \showarticletitle{{LoRa --- A survey of recent research trends}}. In
  \bibinfo{booktitle}{\emph{2018 41st International Convention on Information
  and Communication Technology, Electronics and Microelectronics (MIPRO)}}.
  \bibinfo{publisher}{IEEE}, \bibinfo{address}{Opatija},
  \bibinfo{pages}{0872--0877}.
\newblock
\showISBNx{978-953-233-095-3}
\urldef\tempurl%
\url{https://doi.org/10.23919/MIPRO.2018.8400161}
\showDOI{\tempurl}


\bibitem[\protect\citeauthoryear{SELLER and Sornin}{SELLER and Sornin}{2016}]%
        {seller2016low}
\bibfield{author}{\bibinfo{person}{Olivier~BA SELLER} {and}
  \bibinfo{person}{Nicolas Sornin}.} \bibinfo{year}{2016}\natexlab{}.
\newblock \bibinfo{title}{Low power long range transmitter}.
\newblock
\newblock
\newblock
\shownote{US Patent 9,252,834.}


\bibitem[\protect\citeauthoryear{Seller and Sornin}{Seller and Sornin}{2018}]%
        {seller2018low}
\bibfield{author}{\bibinfo{person}{Olivier Bernard~Andr{\'e} Seller} {and}
  \bibinfo{person}{Nicolas Sornin}.} \bibinfo{year}{2018}\natexlab{}.
\newblock \bibinfo{title}{Low complexity, low power and long range radio
  receiver}.
\newblock
\newblock
\newblock
\shownote{US Patent App. 15/620,364.}


\bibitem[\protect\citeauthoryear{Sinha, Wei, and Hwang}{Sinha
  et~al\mbox{.}}{2017}]%
        {sinha2017survey}
\bibfield{author}{\bibinfo{person}{Rashmi~Sharan Sinha},
  \bibinfo{person}{Yiqiao Wei}, {and} \bibinfo{person}{Seung-Hoon Hwang}.}
  \bibinfo{year}{2017}\natexlab{}.
\newblock \showarticletitle{A survey on LPWA technology: LoRa and NB-IoT}.
\newblock \bibinfo{journal}{\emph{Ict Express}} \bibinfo{volume}{3},
  \bibinfo{number}{1} (\bibinfo{year}{2017}), \bibinfo{pages}{14--21}.
\newblock


\bibitem[\protect\citeauthoryear{Talla, Hessar, Kellogg, Najafi, Smith, and
  Gollakota}{Talla et~al\mbox{.}}{2017}]%
        {Talla2017}
\bibfield{author}{\bibinfo{person}{Vamsi Talla}, \bibinfo{person}{Mehrdad
  Hessar}, \bibinfo{person}{Bryce Kellogg}, \bibinfo{person}{Ali Najafi},
  \bibinfo{person}{Joshua~R. Smith}, {and} \bibinfo{person}{Shyamnath
  Gollakota}.} \bibinfo{year}{2017}\natexlab{}.
\newblock \showarticletitle{{LoRa Backscatter: Enabling The Vision of
  Ubiquitous Connectivity}}.
\newblock \bibinfo{journal}{\emph{Proceedings of the ACM on Interactive,
  Mobile, Wearable and Ubiquitous Technologies}} \bibinfo{volume}{1},
  \bibinfo{number}{3} (\bibinfo{year}{2017}), \bibinfo{pages}{105}.
\newblock
\showISSN{24749567}
\urldef\tempurl%
\url{https://doi.org/10.1145/3130970}
\showDOI{\tempurl}
\showeprint[arxiv]{1705.05953}


\bibitem[\protect\citeauthoryear{Tan, Liu, Fang, Wang, Zhang, Chen, and
  Voelker}{Tan et~al\mbox{.}}{2009}]%
        {Tan2009}
\bibfield{author}{\bibinfo{person}{Kun Tan}, \bibinfo{person}{He Liu},
  \bibinfo{person}{Ji Fang}, \bibinfo{person}{Wei Wang},
  \bibinfo{person}{Jiansong Zhang}, \bibinfo{person}{Mi Chen}, {and}
  \bibinfo{person}{Geoffrey~M. Voelker}.} \bibinfo{year}{2009}\natexlab{}.
\newblock \showarticletitle{{SAM: Enabling Practical Spatial Multiple Access in
  Wireless LAN}}. In \bibinfo{booktitle}{\emph{Proceedings of the 15th annual
  international conference on Mobile computing and networking - MobiCom '09}}
  \emph{(\bibinfo{series}{MobiCom '09})}. \bibinfo{publisher}{ACM},
  \bibinfo{address}{New York, NY, USA}, \bibinfo{pages}{49}.
\newblock
\showISBNx{9781605587028}
\urldef\tempurl%
\url{https://doi.org/10.1145/1614320.1614327}
\showDOI{\tempurl}


\bibitem[\protect\citeauthoryear{Vangelista}{Vangelista}{2017}]%
        {Vangelista2017}
\bibfield{author}{\bibinfo{person}{Lorenzo Vangelista}.}
  \bibinfo{year}{2017}\natexlab{}.
\newblock \showarticletitle{{Frequency Shift Chirp Modulation: The LoRa
  Modulation}}.
\newblock \bibinfo{journal}{\emph{IEEE Signal Processing Letters}}
  \bibinfo{volume}{24}, \bibinfo{number}{12} (\bibinfo{date}{dec}
  \bibinfo{year}{2017}), \bibinfo{pages}{1818--1821}.
\newblock
\showISSN{10709908}
\urldef\tempurl%
\url{https://doi.org/10.1109/LSP.2017.2762960}
\showDOI{\tempurl}


\bibitem[\protect\citeauthoryear{Vangelista, Zanella, and Zorzi}{Vangelista
  et~al\mbox{.}}{2015}]%
        {Vangelista2015}
\bibfield{author}{\bibinfo{person}{Lorenzo Vangelista}, \bibinfo{person}{Andrea
  Zanella}, {and} \bibinfo{person}{Michele Zorzi}.}
  \bibinfo{year}{2015}\natexlab{}.
\newblock \showarticletitle{{Long-range IoT technologies: The dawn of LoRaTM}}.
  In \bibinfo{booktitle}{\emph{Lecture Notes of the Institute for Computer
  Sciences, Social-Informatics and Telecommunications Engineering, LNICST}},
  Vol.~\bibinfo{volume}{159}. \bibinfo{publisher}{Springer, Cham},
  \bibinfo{pages}{51--58}.
\newblock
\showISBNx{9783319270715}
\showISSN{18678211}


\bibitem[\protect\citeauthoryear{Wang, Chen, and Shen}{Wang
  et~al\mbox{.}}{2015}]%
        {wang2015compressive}
\bibfield{author}{\bibinfo{person}{Yingjie Wang}, \bibinfo{person}{Zhiyong
  Chen}, {and} \bibinfo{person}{Manyuan Shen}.}
  \bibinfo{year}{2015}\natexlab{}.
\newblock \showarticletitle{Compressive sensing for uplink cloud radio access
  network with limited backhaul capacity}. In \bibinfo{booktitle}{\emph{2015
  4th International Conference on Computer Science and Network Technology
  (ICCSNT)}}, Vol.~\bibinfo{volume}{1}. IEEE, \bibinfo{pages}{898--902}.
\newblock


\bibitem[\protect\citeauthoryear{{Wright}, {Yang}, {Ganesh}, {Sastry}, and
  {Ma}}{{Wright} et~al\mbox{.}}{2009}]%
        {wright2009}
\bibfield{author}{\bibinfo{person}{J. {Wright}}, \bibinfo{person}{A.~Y.
  {Yang}}, \bibinfo{person}{A. {Ganesh}}, \bibinfo{person}{S.~S. {Sastry}},
  {and} \bibinfo{person}{Y. {Ma}}.} \bibinfo{year}{2009}\natexlab{}.
\newblock \showarticletitle{Robust Face Recognition via Sparse Representation}.
\newblock \bibinfo{journal}{\emph{IEEE Transactions on Pattern Analysis and
  Machine Intelligence}} \bibinfo{volume}{31}, \bibinfo{number}{2}
  (\bibinfo{date}{Feb} \bibinfo{year}{2009}), \bibinfo{pages}{210--227}.
\newblock
\showISSN{0162-8828}
\urldef\tempurl%
\url{https://doi.org/10.1109/TPAMI.2008.79}
\showDOI{\tempurl}


\bibitem[\protect\citeauthoryear{W{\"u}bben, Rost, Bartelt, Lalam, Savin,
  Gorgoglione, Dekorsy, and Fettweis}{W{\"u}bben et~al\mbox{.}}{2014}]%
        {Wubben2014}
\bibfield{author}{\bibinfo{person}{Dirk W{\"u}bben}, \bibinfo{person}{Peter
  Rost}, \bibinfo{person}{Jens~Steven Bartelt}, \bibinfo{person}{Massinissa
  Lalam}, \bibinfo{person}{Valentin Savin}, \bibinfo{person}{Matteo
  Gorgoglione}, \bibinfo{person}{Armin Dekorsy}, {and} \bibinfo{person}{Gerhard
  Fettweis}.} \bibinfo{year}{2014}\natexlab{}.
\newblock \showarticletitle{{Benefits and impact of cloud computing on 5g
  signal processing: Flexible centralization through cloud-RAN}}.
\newblock \bibinfo{journal}{\emph{IEEE Signal Processing Magazine}}
  \bibinfo{volume}{31}, \bibinfo{number}{6} (\bibinfo{date}{nov}
  \bibinfo{year}{2014}), \bibinfo{pages}{35--44}.
\newblock
\showISBNx{1053-5888}
\showISSN{10535888}
\urldef\tempurl%
\url{https://doi.org/10.1109/MSP.2014.2334952}
\showDOI{\tempurl}


\bibitem[\protect\citeauthoryear{Xia, Zhang, Quek, Jin, and Zhu}{Xia
  et~al\mbox{.}}{2018}]%
        {Xia2018}
\bibfield{author}{\bibinfo{person}{Wenchao Xia}, \bibinfo{person}{Jun Zhang},
  \bibinfo{person}{Tony~Q.S. Quek}, \bibinfo{person}{Shi Jin}, {and}
  \bibinfo{person}{Hongbo Zhu}.} \bibinfo{year}{2018}\natexlab{}.
\newblock \showarticletitle{{Joint optimization of fronthaul compression and
  bandwidth allocation in uplink H-CRAN with large system analysis}}.
\newblock \bibinfo{journal}{\emph{IEEE Transactions on Communications}}
  \bibinfo{volume}{66}, \bibinfo{number}{12} (\bibinfo{year}{2018}),
  \bibinfo{pages}{6556--6569}.
\newblock
\showISBNx{9781509050192}
\showISSN{15580857}
\urldef\tempurl%
\url{https://doi.org/10.1109/TCOMM.2018.2861407}
\showDOI{\tempurl}


\bibitem[\protect\citeauthoryear{Xie and Zhang}{Xie and Zhang}{2014}]%
        {Xie2014}
\bibfield{author}{\bibinfo{person}{Xiufeng Xie} {and} \bibinfo{person}{Xinyu
  Zhang}.} \bibinfo{year}{2014}\natexlab{}.
\newblock \showarticletitle{{Scalable user selection for MU-MIMO networks}}. In
  \bibinfo{booktitle}{\emph{Proceedings - IEEE INFOCOM}}.
  \bibinfo{pages}{808--816}.
\newblock
\showISBNx{9781479933600}
\showISSN{0743166X}
\urldef\tempurl%
\url{https://doi.org/10.1109/INFOCOM.2014.6848008}
\showDOI{\tempurl}


\bibitem[\protect\citeauthoryear{Xu, Kim, Huang, Kanhere, Jha, and Hu}{Xu
  et~al\mbox{.}}{2019}]%
        {Xu2019}
\bibfield{author}{\bibinfo{person}{Weitao Xu}, \bibinfo{person}{Jun~Young Kim},
  \bibinfo{person}{Walter Huang}, \bibinfo{person}{Salil~S. Kanhere},
  \bibinfo{person}{Sanjay~K. Jha}, {and} \bibinfo{person}{Wen Hu}.}
  \bibinfo{year}{2019}\natexlab{}.
\newblock \showarticletitle{{Measurement, Characterization, and Modeling of
  LoRa Technology in Multifloor Buildings}}.
\newblock \bibinfo{journal}{\emph{IEEE Internet of Things Journal}}
  \bibinfo{volume}{7}, \bibinfo{number}{1} (\bibinfo{year}{2019}),
  \bibinfo{pages}{298--310}.
\newblock
\showISSN{23274662}
\urldef\tempurl%
\url{https://doi.org/10.1109/jiot.2019.2946900}
\showDOI{\tempurl}
\showeprint[arxiv]{1909.03900}


\end{thebibliography}

% \appendix

\end{document}